\documentclass[aps,prb,reprint,notitlepage,noeprint,superscriptaddress]{revtex4-2}

\usepackage[english]{babel}
\usepackage{amsmath,amssymb,newtxtext,newtxmath,mathdots,bbm,xcolor,graphicx,comment}
\definecolor{darkblue}{rgb}{0.,0.,0.5}
\usepackage[colorlinks,linkcolor=darkblue,citecolor=darkblue,urlcolor=darkblue]{hyperref}
\usepackage{orcidlink}

\newcommand{\sse}{stochastic Schr\"odinger equation}

\newcommand{\C}{\mathbb{C}}

\newcommand{\e}{\mathrm{e}}
\newcommand{\imag}{\mathrm{i}}
\newcommand{\diff}{\mathrm{d}}
\newcommand{\Diff}{\mathrm{D}}

\newcommand{\ket}[1]{\lvert #1 \rangle}
\newcommand{\bra}[1]{\langle #1 \rvert}
\newcommand{\braket}[1]{\langle #1 \rangle}

\newcommand{\svn}[1]{S_{#1}}

\newcommand{\dQG}{\delta Q_{\mathcal{G}}}
\newcommand{\dQS}{\delta Q_{\Sigma}}

\makeatletter
\newcommand*{\transpose}{%
  {\mathpalette\@transpose{}}%
}
\newcommand*{\@transpose}[2]{%
  \raisebox{\depth}{$\m@th#1\intercal$}%
}
\makeatother

\newcommand{\abs}[1]{\left\lvert #1 \right\rvert}
\newcommand{\norm}[1]{\left\lVert #1 \right\rVert}
\newcommand{\tr}{\mathop{\mathrm{tr}}}
\newcommand{\trK}{\mathop{\mathrm{tr}_K}}
\newcommand{\trR}{\mathop{\mathrm{tr}_R}}

\newcommand{\Tr}{\mathop{\mathrm{Tr}}}
\newcommand{\detK}{\mathop{\mathrm{det}_K}}

\newcommand{\sgn}{\mathop{\mathrm{sgn}}}

\renewcommand{\Re}{\mathop{\mathrm{Re}}}

\makeatletter
\DeclareFontFamily{OMX}{MnSymbolE}{}
\DeclareSymbolFont{MnLargeSymbols}{OMX}{MnSymbolE}{m}{n}
\SetSymbolFont{MnLargeSymbols}{bold}{OMX}{MnSymbolE}{b}{n}
\DeclareFontShape{OMX}{MnSymbolE}{m}{n}{
  <-6>  MnSymbolE5
  <6-7>  MnSymbolE6
  <7-8>  MnSymbolE7
  <8-9>  MnSymbolE8
  <9-10> MnSymbolE9
  <10-12> MnSymbolE10
  <12->   MnSymbolE12
}{}
\DeclareFontShape{OMX}{MnSymbolE}{b}{n}{
  <-6>  MnSymbolE-Bold5
  <6-7>  MnSymbolE-Bold6
  <7-8>  MnSymbolE-Bold7
  <8-9>  MnSymbolE-Bold8
  <9-10> MnSymbolE-Bold9
  <10-12> MnSymbolE-Bold10
  <12->   MnSymbolE-Bold12
}{}

\let\llangle\@undefined
\let\rrangle\@undefined
\DeclareMathDelimiter{\llangle}{\mathopen}%
{MnLargeSymbols}{'164}{MnLargeSymbols}{'164}
\DeclareMathDelimiter{\rrangle}{\mathclose}%
{MnLargeSymbols}{'171}{MnLargeSymbols}{'171}
\makeatother

\begin{document}

\title{Absence of measurement- and unraveling-induced entanglement transitions
  in continuously monitored one-dimensional free fermions}

\author{Clemens Niederegger\orcidlink{0009-0008-4766-3131}}

\affiliation{Institute for Theoretical Physics, University of Innsbruck, 6020
Innsbruck, Austria}

\author{Tatiana Vovk\orcidlink{0000-0001-6369-4079}}

\affiliation{Institute for Theoretical Physics, University of Innsbruck, 6020
  Innsbruck, Austria}

\affiliation{Institute for Quantum Optics and Quantum Information, Austrian
  Academy of Sciences, 6020 Innsbruck, Austria}

\author{Elias Starchl\orcidlink{0000-0002-3357-9173}}

\affiliation{Institute for Theoretical Physics, University of Innsbruck, 6020
Innsbruck, Austria}

\author{Lukas M. Sieberer\orcidlink{0000-0002-0163-7850}}

\email{lukas.sieberer@uibk.ac.at}

\affiliation{Institute for Theoretical Physics, University of Innsbruck, 6020
Innsbruck, Austria}

\begin{abstract}
  Continuous monitoring of one-dimensional free fermionic systems can generate
  phenomena reminiscent of quantum criticality, such as logarithmic entanglement
  growth, algebraic correlations, and emergent conformal invariance, but in a
  nonequilibrium setting. However, whether these signatures reflect a genuine
  phase of nonequilibrium quantum matter or persist only over finite length
  scales is an active area of research. We address this question in a free
  fermionic chain subject to continuous monitoring of lattice-site
  occupations. An unraveling phase $\varphi$ interpolates between measurement
  schemes, corresponding to different stochastic unravelings of the same
  Lindblad master equation: For $\varphi = 0$, measurements disentangle lattice
  sites, while for $\varphi = \pi/2$ they act as unitary random noise, yielding
  volume-law steady-state entanglement. Using replica Keldysh field theory, we
  obtain a nonlinear sigma model describing the long-wavelength physics. This
  analysis shows that for $0 \leq \varphi < \pi/2$, entanglement ultimately
  obeys an area law, but only beyond the exponentially large scale
  $\ln(l_{\varphi, *}) \sim J/[\gamma \cos(\varphi)]$, where $J$ is the hopping
  amplitude and $\gamma$ the measurement rate. Resolving $l_{\varphi, *}$ in
  numerical simulations is difficult for $\gamma/J \to 0$ or
  $\varphi \to \pi/2$. However, the theory also predicts that critical-like
  behavior appears below a crossover scale that grows only algebraically in
  $J/\gamma$, making it numerically accessible. Our simulations confirm these
  predictions, establishing the absence of measurement- or unraveling-induced
  entanglement transitions in this model.
\end{abstract}

\maketitle

\section{Introduction}
\label{sec:introduction}

Continuously monitored free fermions provide a versatile framework for exploring
the interplay between unitary dynamics and measurements in quantum many-body
systems~\cite{Cao2019, Chen2020, Nahum2020, Alberton2021, Turkeshi2021,
  Tang2021, Bao2021, Buchhold2021, Sang2021, Fidkowski2021, VanRegemortel2021,
  Coppola2022, Carollo2022, Ladewig2022, Turkeshi2022, Turkeshi2022a,
  *Turkeshi2023, Buchhold2022, Piccitto2022, Muller2022, Zhang2022, Minato2022,
  Jian2022, Jian2023, Yang2023a, Szyniszewski2023, Szyniszewski24,
  Paviglianiti2023, Loio2023, LeGal2023, Granet2023, Lavasani2023, Feng2023,
  Poboiko2023, Russomanno23, Poboiko2024, Kells2023, Fava2023, Merritt2023,
  Klocke2023, Wang2024, Chahine2024, DiFresco2024, Xiao2024, Turkeshi2024,
  LeGal2024, Liu2024, DiFresco2024a, Lumia2024, Tsitsishvili2024, Fava2024,
  Guo2025, Poboiko2025, Eissler2025, Matsubara2025, Chatterjee2025, Starchl2025,
  Paviglianiti2025, Poboiko2025a, Muller2025, Klocke2025, Li2025,
  Ravindranath2025, Huang2025, Soares2025, Muzzi2025, Leung2025, Chahine2025,
  Qiu2025, Bhuiyan2025, Poboiko2025b, Fan2025}. They display a rich
phenomenology while still being amenable to efficient numerical simulations and
tractable through field-theoretic analytical methods. Nevertheless, fundamental
questions concerning the existence of measurement-induced entanglement
transitions in certain models remain a subject of ongoing investigation.

Here, we study a one-dimensional (1D) lattice system of free fermions subject to
a continuous measurement process which can be regarded as a stochastic
unraveling of a Lindblad master equation. An \emph{unraveling phase} $\varphi$
allows us to tune between different measurement schemes, and thus different
unravelings of the same master equation. Our analytical and numerical results
indicate that in this model, there is no entanglement transition that can be
induced by either changing the measurement rate $\gamma$ or the unraveling phase
$\varphi$. Therefore, our findings suggest that the apparent Kosterlitz-Thouless
(KT) transition to a critical phase, which was observed in numerical simulations
of similar models~\cite{Alberton2021, Eissler2025}, is in fact a finite-size
crossover phenomenon.

In view of the ongoing debate about the existence of a critical phase and a
measurement-induced phase transition in free fermionic systems with a conserved
number of particles, we begin by briefly reviewing previous studies of related
models.

\subsection{Previous work}
\label{sec:previous-work}

The study of entanglement dynamics in continuously monitored free fermionic
systems was initiated in Ref.~\cite{Cao2019}, which considered a 1D
tight-binding chain subject to continuous measurements of lattice-site
occupation numbers. Based on a phenomenological quasiparticle picture, the
authors concluded that no measurement-induced entanglement transition occurs in
this model; rather, the stationary entanglement obeys an area law for any
nonzero measurement rate $\gamma$.

This conclusion was challenged by numerical results from Alberton et
al.~\cite{Alberton2021}, which suggested a KT transition from an area-law phase
to a critical phase at low measurement rates. The critical phase is
characterized by logarithmic growth of the entanglement entropy with subsystem
size, algebraically decaying correlations, and emergent conformal
invariance. These findings were further supported by a field-theoretic analysis
of Dirac fermions~\cite{Buchhold2021}. Subsequently, several modifications of
the model studied in Refs.~\cite{Cao2019, Alberton2021} were found to display
similar phenomenology. These include long-range hopping~\cite{Muller2022,
  Minato2022, Qiu2025}, spatial disorder~\cite{Szyniszewski2023, Szyniszewski24,
  Matsubara2025, Qiu2025, Zhao2025}, Floquet driving~\cite{Chatterjee2025}, as
well as tunable measurement schemes that interpolate continuously between
different unravelings of the same Lindblad master equation~\cite{Eissler2025}.

In the context of tunable measurements, the quantum state diffusion equation
used to model continuous monitoring in Refs.~\cite{Cao2019, Alberton2021} can be
viewed as one particular stochastic unraveling of the open-system dynamics of
the average state, which are governed by a Markovian quantum master equation in
Lindblad form. However, the unraveling of a Lindblad equation into quantum
trajectories is not unique and depends on the specific type of continuous
measurement performed~\cite{Jacobs2006, Gardiner2015, Wiseman2010,
  Jacobs2014}. Continuous monitoring can lead to either quantum jumps or quantum
state diffusion, described by distinct stochastic Schr\"odinger equations. In
Ref.~\cite{Alberton2021}, both types of trajectories were found to yield
qualitatively similar results. However, even when restricting to quantum state
diffusion, the choice of unraveling remains nonunique: Besides the unraveling
considered in Refs.~\cite{Cao2019, Alberton2021}, where entangling unitary
dynamics compete with disentangling measurements, there exists a \emph{unitary
  unraveling} of the same master equation that describes time evolution under
randomly fluctuating onsite potentials~\cite{Cao2019}. For this unitary
unraveling, the steady-state entanglement entropy exhibits volume-law scaling,
raising the question of whether an entanglement transition occurs when
interpolating between the conventional and unitary unravelings, analogous to
unraveling-induced phase transitions observed in quantum
circuits~\cite{Vovk2022, Vovk2024}. This question was answered affirmatively in
Ref.~\cite{Eissler2025}: Numerical results indicate that not only changing the
measurement rate but also tuning the unraveling can induce a KT transition from
an area-law to a critical phase, while volume-law entanglement occurs only at a
single point in parameter space.

A surprising new development emerged from the study of yet another measurement
protocol~\cite{Poboiko2023}: projective measurements of local occupation
numbers, performed at random points in space and time, similar to the type of
measurements commonly considered in hybrid random
circuits~\cite{Fisher2023}. The resulting dynamics are described on large scales
by a nonlinear sigma model (NLSM). Such models are familiar from the theory of
disordered electronic systems~\cite{Evers2008}. The connection between
$1 + 1 \mathrm{D}$ measurement-induced dynamics and the physics of Anderson
localization in 2D disordered systems implies that the logarithmic growth of the
entanglement entropy with subsystem size is cut off at a length scale $l_{*}$
that is exponentially large in the inverse measurement rate,
$\ln(l_{*}) \sim \gamma^{-1}$, corresponding to the crossover from weak to
strong localization in 2D systems. This crossover occurs for any nonzero
measurement rate, indicating that random projective measurements do not induce
an entanglement transition. Nevertheless, the exponential dependence of $l_{*}$
on $\gamma$ makes the crossover from logarithmic to area-law entanglement appear
sharp, which can be misinterpreted as a phase transition in numerical studies. A
direct numerical verification of this dependence has only recently been
achieved~\cite{Fan2025}.

As stated above, the measurement processes considered in
Refs.~\cite{Alberton2021} and~\cite{Poboiko2023} are different: quantum state
diffusion and quantum jump trajectories in the former, and random projective
measurements in the latter. However, the studied models have the same
symmetries, spatial dimensionality, and are restricted to short-range
hopping. Therefore, the same large-scale behavior described by the NLSM is
expected to apply universally to all of them. This expectation is corroborated
by the results of Refs.~\cite{Fava2024, Starchl2025}, which derived the same
NLSM for yet other types of Hamiltonian dynamics and measurements and clarified
the underlying symmetry requirements, in terms of particle-hole
symmetry~\cite{Fava2024} and the role of particle-number
conservation~\cite{Starchl2025}. Numerical evidence for the absence of an
entanglement transition is also presented in Refs.~\cite{Coppola2022, Kells2023,
  Fan2025}. Notably, the description in terms of an NLSM does predict an
entanglement transition in higher dimensions~\cite{Chahine2024, Poboiko2024} or
if particle-number conservation is broken~\cite{Fava2023}.

The developments outlined above---the accumulation of evidence for a KT
transition to a critical phase on the one hand, and for the instability of the
critical phase in the thermodynamic limit on the other hand---have progressed in
parallel in recent years. By now, the instability of the critical phase has been
corroborated by numerical results for specific models. However, is it possible
to stabilize the critical phase through one of the modifications of these models
mentioned above, such as spatial disorder, Floquet driving, and different
unravelings?  Addressing that question analytically within the replica Keldysh
field theory framework that underlies the NLSM description is in general not
straightforward. Demonstrating the instability of the critical phase numerically
is equally challenging: Since the scale $l_{*}$, beyond which we should expect
area-law scaling of the entanglement entropy, is exponentially large in the
inverse measurement rate, its direct observation is extremely
demanding~\cite{Fan2025}. An alternative manifestation of the NLSM physics is
the logarithmically slow RG flow responsible for the dependence
$\ln(l_{*}) \sim \gamma^{-1}$. This flow is reflected in the weak-localization
correction~\cite{Poboiko2023, Poboiko2025}. However, to determine the
weak-localization correction, detailed knowledge of the Gaussian theory that
describes the logarithmic phase in the absence of RG corrections is required.

Signatures of the instability of the critical phase that are accessible more
directly and on only algebraically large scales were proposed in
Ref.~\cite{Starchl2025}. These are based on the analytical result, derived from
the NLSM description, that features characteristic of a critical phase are
confined to a finite critical range of length scales, bounded from below by
$l_0 \sim \gamma^{-1}$ and from above by $l_c \sim \gamma^{-2}$. In particular,
the connected density correlation function shows algebraic decay only up to a
crossover scale $l_c \sim \gamma^{-2}$. Moreover, the scale-dependent effective
central charge, defined as
$c_{\ell} = 3 \partial \svn{\ell}/\partial \! \ln(\ell)$, where $S_{\ell}$ is
the von Neumann entanglement entropy of a subsystem of size $\ell$, has a
maximum at $l_m \sim \gamma^{-3/2}$, between $l_0$ and $l_c$. Only in the
vicinity of this maximum is the growth of the entanglement entropy approximately
logarithmic. The decrease of the effective central charge on scales greater than
$l_m$ is a clear indication of slow crossover toward area-law
scaling. Importantly, the scaling of $l_c$ and $l_m$ with $\gamma$ is observable
over a wide range of values of the measurement rate, including across the
purported critical point~\cite{Starchl2025}. These and further signatures of
smooth scaling with $\gamma$, in agreement with predictions from the NLSM,
provide strong numerical support for the absence a measurement-induced
entanglement transition in the models studied in Ref.~\cite{Starchl2025}.

\subsection{Synopsis}

In this paper, we use the same ideas to demonstrate the absence of measurement-
and unraveling-induced entanglement transitions in free fermions on a 1D lattice
under continuous measurements of lattice-site occupation numbers. Similarly to
the model of Ref.~\cite{Eissler2025}, our model features an unraveling phase
$\varphi$ that allows us to interpolate between conventional quantum state
diffusion~\cite{Cao2019, Alberton2021} for $\varphi = 0$ and the unitary
unraveling~\cite{Cao2019} for $\varphi = \pi/2$. For any value of
$0 \leq \varphi < \pi/2$, the model is described on large scales by the same
NLSM as was found previously in Ref.~\cite{Chahine2024} for quantum state
diffusion and in Refs.~\cite{Poboiko2023, Poboiko2024, Poboiko2025, Fava2024,
  Starchl2025} for other types of dynamics or measurements, or both. The NLSM
description implies that in one spatial dimension, there is no entanglement
transition that can be induced by changing either the measurement rate $\gamma$
or the unraveling phase $\varphi$. On large scales, the entanglement entropy
always obeys area-law scaling. This indicates that the KT transition to a
critical phase observed in Refs.~\cite{Alberton2021, Eissler2025} is, in fact, a
finite-size crossover phenomenon. We validate the analytical predictions
obtained from the NLSM description by determining the weak-localization
correction numerically, finding good agreement with the universal forms for
intact and broken particle-hole symmetry when $\varphi = 0$ and
$\varphi \neq 0$, respectively. Furthermore, we demonstrate the scaling
$l_c \sim \gamma^{-2}$ and $l_m \sim \gamma^{-3/2}$ for values of the
unravelling phase $\varphi$ below a specific value. The expected scaling is
masked by nonuniversal modifications on short scales for larger values of
$\varphi$.

We find that the effect of increasing $\varphi$ at fixed $\gamma$ is
qualitatively similar to \emph{decreasing} $\gamma$ at fixed $\varphi$, which is
compatible with the observation of an apparent unraveling-induced transition in
Ref.~\cite{Eissler2025}. For $\varphi \to \pi/2$, the scale $l_{\varphi, *}$
which marks the onset of area-law entanglement diverges,
$l_{\varphi, *} \to \infty$. Exactly at $\varphi = \pi/2$, since the unitary
unraveling corresponds to classical noise, the system exhibits volume-law
entanglement for any value of the measurement rate.

The paper is organized as follows. In Sec.~\ref{sec:model}, we introduce our
model and discuss the physical significance of the unraveling phase
$\varphi$. We present the analytical description of our model using replica
Keldysh field theory in Sec.~\ref{sec:repl-keldysh-field}. In particular, we
discuss how our derivation differs from previous ones and how it allows us to
classify the resulting Keldysh action in terms of its symmetries. We analyze the
field theory within the Gaussian approximation in
Sec.~\ref{sec:gaussian-theory}. The Gaussian theory describes a critical phase
with algebraic decay of correlations and logarithmic growth of the stationary
entanglement entropy. In Sec.~\ref{sec:NLSM}, we briefly summarize the
long-wavelength description in terms of an NLSM, focusing on the case of broken
particle-hole symmetry for $0 < \varphi < \pi/2$. We validate the analytical
predictions, obtained by incorporating the RG flow of the NLSM into the Gaussian
theory, in Sec.~\ref{sec:numerics-vs-theroy} through a detailed comparison with
numerical simulations. Conclusions are presented in
Sec.~\ref{sec:conclusions}. Two appendices contain technical details of our
analytical and numerical studies.

\section{Model}
\label{sec:model}

We consider free fermions on a 1D lattice that are subject to continuous
measurements. The Hamiltonian reads
\begin{equation}
  \label{eq:Hamiltonian}
  \hat{H} = - J \sum_{l = 1}^L \left( \hat{\psi}_l^{\dagger} \hat{\psi}_{l +
      1}^{} + \hat{\psi}_{l + 1}^{\dagger} \hat{\psi}_l^{} \right),
\end{equation}
where $J$ is the hopping amplitude, $L$ is the length of the 1D chain, and
$\hat{\psi}_l$ and $\hat{\psi}_l^{\dagger}$ are fermionic annihilation and
creation operators, respectively. Periodic boundary conditions are implemented
by setting $\hat{\psi}_{L + 1} = \hat{\psi}_1$. The time evolution of the system
under continuous monitoring is described by a stochastic Schr\"odinger
equation~\cite{Jacobs2006, Gardiner2015, Wiseman2010, Jacobs2014}
\begin{multline}
  \label{eq:sse}
  \ket{\psi(t + \diff t)} = \left\{ 1 - \imag \hat{H} \diff t - \frac{\gamma}{2}
    \sum_{l = 1}^L \left[ \hat{n}_l - \langle \hat{n}_l(t) \rangle \right]^2
    \diff t \right. \\ \left. + \sqrt{\gamma} \sum_{l = 1}^L \left[ \hat{n}_l -
      \langle \hat{n}_l(t) \rangle \right] \e^{\imag \varphi} \diff W_l(t)
  \right\} \ket{\psi(t)}.
\end{multline}
Here, $\ket{\psi(t)}$ is the conditional state that depends on the measurement
record, and
$\langle \hat{n}_l(t) \rangle = \braket{\psi(t) | \hat{n}_l | \psi(t)}$ is the
expectation value of $\hat{n}_l = \hat{\psi}_l^{\dagger} \hat{\psi}_l^{}$ in the
conditional state. The randomness of outcomes, which is inherent to quantum
measurements, is described by the real stochastic Wiener increments
$\diff W_l(t)$. These are independent for different lattice sites
$l \in \{ 1, \dotsc, L \}$ and obey the conditions
\begin{equation}
  \label{eq:Ito-rules}
  \overline{\diff W_l(t)} = 0, \qquad \diff W_l(t) \diff W_{l'}(t) = \delta_{l,
    l'} \diff t,
\end{equation}
where the overline denotes the average over quantum trajectories corresponding
to different measurement outcomes. In Eq.~\eqref{eq:sse}, the Wiener increment
is multiplied by a phase factor that contains the unraveling phase
$\varphi$. The physical meaning of the unraveling phase becomes apparent from a
formal derivation of the stochastic Schr\"odinger equation, which we sketch in
the following and discuss in more detail in Appendix~\ref{sec:sse-derivation}.

The minimal setting required to introduce the unraveling phase $\varphi$ is a
single lattice site $l$ coupled to an ancilla bosonic mode, which is initially
in the vacuum state of the bosonic annihilation operator $\hat{b}$. In a
discrete time step of duration $\Delta t$, the quantum states of the lattice
site and the bosonic ancilla are entangled through the application of the
unitary
$\hat{U} = \e^{\sqrt{\gamma} \hat{n}_l \left( \hat{b}^{\dagger} - \hat{b}
  \right)}$.
In other words, information about $\hat{n}_l$ is imprinted on the ancilla
through a displacement of the bosonic vacuum. This information is then read out
by measuring the phase-dependent field quadrature
$\hat{x}_{\varphi} = \bigl( \hat{b} \e^{\imag \varphi} + \hat{b}^{\dagger} \e^{-
  \imag \varphi} \bigr) / \sqrt{2}$.
The stochastic Schr\"odinger equation~\eqref{eq:sse} is obtained by modeling the
measurement outcomes as a classical stochastic process, generalizing the
dynamics to an extended lattice system, and taking the limit of infinitesimal
time steps, $\Delta t \to \diff t$. While we focus here on reading out
information about $\hat{n}_l$, the derivation of Eq.~\eqref{eq:sse} remains
valid for any Hermitian system observable.

The above discussion clarifies the meaning of the unraveling phase $\varphi$ in
a formal measurement performed on an ancilla bosonic mode. How could such a
measurement be implemented in practice? A natural physical realization of the
model described by Eq.~\eqref{eq:sse} is provided by fermionic atoms trapped in
an optical lattice, where the ancillary bosonic mode is represented by an
optical field. For example, a continuous measurement of the atomic density can
be realized by dispersively coupling the density to the light field of a driven
optical cavity and performing homodyne detection on the light leaking out of the
cavity~\cite{Yang2018}. In homodyne detection, the cavity output field is
interfered with a coherent phase reference, the local oscillator. The phase of
this reference field corresponds precisely to the unraveling phase
$\varphi$~\cite{Gardiner2015}.

Interestingly, the unconditional dynamics of the density matrix
$\hat{\rho}(t) = \overline{\ket{\psi(t)} \bra{\psi(t)}}$, obtained as the
average of the state $\ket{\psi(t)}$ over the ensemble of trajectories, do not
depend on the choice of $\varphi$ and are described by the quantum master
equation for dephasing,
\begin{equation}
  \label{eq:meq}
  \frac{\diff}{\diff t} \hat{\rho}(t) = - \imag \left[ \hat{H}, \hat{\rho}(t)
  \right] - \frac{\gamma}{2} \sum_{l = 1}^L \left[ \hat{n}_l, \left[ \hat{n}_l,
      \hat{\rho}(t) \right] \right].
\end{equation}
The stochastic Schr\"odinger equation~\eqref{eq:sse} can be regarded as an
unraveling of the master equation~\eqref{eq:meq}. From this perspective,
different choices of $\varphi$ correspond to different unravelings of the same
master equation~\cite{Cao2019, Vovk2022, Vovk2024, Eissler2025}. For this
reason, we refer to $\varphi$ as the unraveling phase.

The unraveling phase $\varphi$ allows us to interpolate between conventional
quantum state diffusion~\cite{Cao2019, Alberton2021} for $\varphi = 0$ and the
unitary unraveling corresponding to randomly fluctuating onsite potentials for
$\varphi = \pi/2$~\cite{Cao2019}. To show this, we perform a time-dependent
gauge transformation~\cite{Gardiner2015},
\begin{equation}
  \label{eq:gauge-transformation}
  \begin{split}
    \ket{\psi(t)} & \mapsto \e^{- \imag \phi(t)} \ket{\psi(t)}, \\ \diff \phi(t)
    & = \sqrt{\gamma} \sin(\varphi) \sum_{l = 1}^L \langle \hat{n}_l(t) \rangle
    \diff W_l(t).
  \end{split}
\end{equation}
Then, the stochastic Schr\"odinger equation~\eqref{eq:sse} is recast as
\begin{multline}
  \ket{\psi(t + \diff t)} = \left( 1 - \imag \hat{H} \diff t - \frac{\gamma}{2}
    \sum_{l = 1}^L \left\{ \cos(\varphi)^2 \left[ \hat{n}_l - \langle
        \hat{n}_l(t) \rangle \right]^2 \right. \right. \\ \left. +
    \sin(\varphi)^2 \hat{n}_l \right\} \diff t + \sqrt{\gamma} \sum_{l = 1}^L
  \left\{ \cos(\varphi) \left[ \hat{n}_l - \langle \hat{n}_l(t) \rangle \right]
  \right. \\ \left. \vphantom{\sum_{l = 1}^L} \left. + \imag \sin(\varphi)
      \hat{n}_l \right\} \diff W_l(t) \right) \ket{\psi(t)}.
\end{multline}
The terms involving $\cos(\varphi)$ describe continuous monitoring of the field
quadrature
$\hat{x}_{\varphi = 0} = \bigl( \hat{b} + \hat{b}^{\dagger} \bigr) / \sqrt{2}$,
but with an effective measurement rate $\gamma \cos(\varphi)^2$. In contrast,
the terms involving $\sin(\varphi)$ can be regarded as describing onsite
potentials and thus add temporally fluctuating disorder to the unitary dynamics
governed by the Hamiltonian~\cite{Cao2019, Eissler2025}. From this perspective,
by increasing the unraveling phase from $\varphi = 0$ to $\varphi = \pi/2$, we
interpolate continuously between measurements of local occupation numbers and
classical noise~\cite{Gardiner2014}. The effect of classical noise is opposite
to that of measurements: While local measurements disentangle the quantum state,
noise induces heating and thus leads to the growth of entanglement. Note that
this observation is not affected by the gauge
transformation~\eqref{eq:gauge-transformation} and remains valid in the original
frame. We thus expect increasing $\varphi$ to have a similar effect as
decreasing the measurement rate $\gamma$~\cite{Vovk2022}.

\section{Replica Keldysh field theory}
\label{sec:repl-keldysh-field}

Our goal is to understand the dynamics described by Eq.~\eqref{eq:sse} and the
dependence on the parameters $\gamma$ and $\varphi$. To this end, we first
develop an analytical approach in the framework of replica Keldysh field
theory~\cite{Fava2023, Jian2023, Poboiko2023, Poboiko2024, Fava2024, Guo2025,
Chahine2024, Poboiko2025}. We later confirm the analytical findings through
direct numerical simulations of Eq.~\eqref{eq:sse}.

A reformulation of Eq.~\eqref{eq:sse} with $\varphi = 0$ as a fermionic replica
Keldysh field theory was derived previously by Chahine and
Buchhold~\cite{Chahine2024}. In their approach, the average over measurement
outcomes is taken for the $R$-replica stochastic master equation, and the
resulting deterministic master equation is reformulated as a Keldysh functional
integral. Here instead, we construct the Keldysh functional integral from the
linear stochastic master equation for the unnormalized density matrix; the
resulting Keldysh partition function, which still depends on the measurement
outcomes, is quadratic in fermionic fields. This form allows for a particularly
transparent analysis of symmetries. In particular, we find that for
$\varphi = 0$, the theory has a particle-hole symmetry~\cite{Fava2023,
  Poboiko2025, Starchl2025}, which is broken for $\varphi \neq 0$. The average
over trajectories is taken as a final step.

\subsection{Linear stochastic master equation and replica trick}
\label{sec:line-stoch-mast}

In deriving a functional-integral description for the dynamics described by
Eq.~\eqref{eq:sse}, we have to deal with two technical issues: (i)~The
dependence on the expectation values $\langle \hat{n}_l(t) \rangle$ renders the
stochastic Schr\"odinger equation~\eqref{eq:sse} nonlinear in the conditional
state $\ket{\psi(t)}$. This applies also to the corresponding stochastic master
equation, which is therefore not a suitable starting point for constructing a
functional integral description of the dynamics---the resulting action would
depend, via the expectation value $\langle \hat{n}_l(t) \rangle$, on the
functional integral itself. (ii)~Not only the dynamics, but also the observables
of interest are nonlinear in the conditional state. Indeed, all linear
observables can be calculated from the density matrix $\hat{\rho}(t)$ that obeys
the master equation~\eqref{eq:meq}. The steady state of the latter is a fully
mixed state at infinite temperature, irrespective of the value of
$\gamma$. Therefore, nontrivial effects of measurements manifest only in
nonlinear observables. Important examples, which we focus on in our work and
introduce further below, are the connected density correlation function and the
entanglement entropy~\cite{Cao2019, Alberton2021, Poboiko2023, Starchl2025}.

To overcome (i), we propose to construct the functional integral starting from
the linear evolution equation for the unnormalized density matrix $\hat{D}(t)$,
which we derive in Appendix~\ref{sec:sse-derivation}. With the construction of
the functional integral in mind, we discretize the evolution of $\hat{D}(t)$
from $t_0$ to $t$ into $N$ time steps of duration $\Delta t = (t - t_0)/N$.  The
linear evolution equation for the unnormalized density matrix at discrete times,
$\hat{D}_n = \hat{D}(t_n)$ for $t_n = t_0 + n \Delta t$, reads
\begin{multline}
  \label{eq:linear-sme}
  \hat{D}_{n + 1} = \hat{D}_n - \imag \left[ \hat{H}, \hat{D}_n \right] \Delta t
  - \frac{\gamma}{2} \sum_{l = 1}^L \left[ \hat{n}_l, \left[ \hat{n}_l,
      \hat{D}_n \right] \right] \Delta t \\ + \sqrt{\gamma} \sum_{l = 1}^L
  \left( \hat{n}_l \e^{\imag \varphi} \hat{D}_n + \hat{D}_n \hat{n}_l \e^{-
      \imag \varphi} \right) \Delta W_{l, n}.
\end{multline}
As the derivation presented in Appendix~\ref{sec:sse-derivation} shows, this
equation is equivalent to a linear evolution equation for the pure state
$\ket{v_n} = \ket{v(t_n)}$. Therefore, Eq.~\eqref{eq:linear-sme} preserves the
purity of the state $\hat{D}_n = \ket{v_n} \bra{v_n}$; however, it does not
preserve the norm of the state. According to Born's rule, the norm encodes the
probability for a trajectory associated with a sequence of stochastic
increments,
\begin{equation}
  \Delta \mathbf{W} = \left( \Delta W_{l, n} \right)_{l \in \{ 1, \dotsc, L\}, n
    \in \{ 0, \dotsc, N - 1 \}},
\end{equation}
to occur. More precisely, the probability of a trajectory is
the product of the ostensible distribution of
$\Delta \mathbf{W}$~\cite{Wiseman2010}, given by
\begin{equation}
  \label{eq:Q-Delta-W}
  Q_{\Delta \mathbf{W}} = \frac{1}{\left( 2 \pi \Delta t
    \right)^{N L/2}} \exp \! \left( - \frac{1}{2 \Delta t} \sum_{l = 1}^L
    \sum_{n = 0}^{N - 1} \Delta W_{l, n}^2 \right),
\end{equation}
and the norm of the state, $\norm{\ket{v_N}} = \tr \! \left( \hat{D}_N \right)$
at time $t_N = t$.

Having stated the linear evolution equation~\eqref{eq:linear-sme} that serves as
the starting point for deriving a functional integral description, we now turn
to issue (ii) identified above: the question of how to evaluate trajectory
averages of nonlinear observables. As we discuss next, this can be achieved by
introducing replicas.

The product of the expectation values of two operators $\hat{A}$ and $\hat{B}$
at time $t_N$ provides a basic example of a nonlinear observable. To evaluate
the average over trajectories, we introduce $R$ replicas of the system,
$\hat{D}_{r, n}$, with replica index $r \in \{ 1, \dotsc, R \}$. We can thus
write
\begin{multline}  
    \overline{\langle \hat{A}_N \rangle \langle \hat{B}_N \rangle} = \int
    \diff \mu_{\Delta \mathbf{W}} \tr \! \left( \hat{D}_N \right)^{-1} \tr \!
    \left( \hat{A} \hat{D}_N \right) \tr \! \left( \hat{B} \hat{D}_N \right) \\
    \begin{aligned}
      & = \lim_{R \to 1} \int \diff \mu_{\Delta \mathbf{W}} \tr \! \left(
      \hat{D}_N \right)^{R - 2} \tr \!
    \left( \hat{A} \hat{D}_N \right) \tr \! \left( \hat{B} \hat{D}_N \right) \\
    & = \lim_{R \to 1} \int \diff \mu_{\Delta \mathbf{W}} \tr \! \left[ \left(
        \hat{A}_1 \otimes \hat{B}_2 \right) \left( \bigotimes_{r = 1}^R
        \hat{D}_{r, N} \right) \right],
  \end{aligned}
\end{multline}
where
$\diff \mu_{\Delta \mathbf{W}} = \diff \Delta \mathbf{W} \, Q_{\Delta
  \mathbf{W}}$.
The assumption underlying the replica trick is that the last equality still
holds in the limit $R \to 1$ even though the formal manipulation of the
integrand is valid only for integer $R > 2$~\cite{Altland2010a}.

Generalizing the above to a product of more than two expectation values is
straightforward. To calculate any such average, it is convenient to introduce
appropriate sources in the Hamiltonian so as to turn the Keldysh partition
function for $R$ replicas,
\begin{equation}
  \label{eq:Z-R}
  Z_R = \int \diff \mu_{\Delta \mathbf{W}} \tr \! \left[ \bigotimes_{r = 1}^R
    \hat{D}_{r, N} \right]  = \int \diff \mu_{\Delta \mathbf{W}} \prod_{r = 1}^R
  \tr \! \left[ \hat{D}_r(t) \right],
\end{equation}
into a generating functional. Crucially, the right-hand side of
Eq.~\eqref{eq:Z-R} is linear in the unnormalized density matrix of $R$
replicas. Therefore, a functional integral representation of the Keldysh
partition function can be obtained via the usual Keldysh
construction~\cite{Kamenev2023, Altland2010a, Sieberer2016a, Sieberer2025},
which we summarize below.

\subsection{Replica Keldysh action}
\label{sec:replica-keldysh-action}

According to Eq.~\eqref{eq:Z-R}, before we take the average over
$\Delta \mathbf{W}$, the Keldysh partition function factorizes over
replicas. Therefore, we consider first a single replica and write
Eq.~\eqref{eq:linear-sme} as
\begin{equation}
  \label{eq:replica-linear-sme}
  \hat{D}_{r, n + 1} = \left( 1 + \mathcal{L}_n \right) \hat{D}_{r, n},
\end{equation}
where
\begin{multline}
  \label{eq:L-n}
  \mathcal{L}_n \hat{D}_{r, n} = - \imag \left[ \hat{H}, \hat{D}_{r, n} \right] \Delta t
  + \frac{\gamma}{2} \sum_{l = 1}^L \left[ \hat{n}_l, \left[ \hat{n}_l,
      \hat{D}_{r, n} \right] \right] \Delta t \\ + \sqrt{\gamma} \sum_{l = 1}^L
  \left( \hat{n}_l \e^{\imag \varphi} \hat{D}_{r, n} + \hat{D}_{r, n} \hat{n}_l \e^{-
      \imag \varphi} \right) \Delta W_{l, n}.
\end{multline}
Next, we represent $\hat{D}_{r, n}$ in a basis of fermionic coherent states:
\begin{multline}
  \label{eq:coherent-state-representation}
  \hat{D}_{r, n} = \int \diff \psi_{+, r, n}^{*} \, \diff \psi_{+, r, n} \,
  \diff \psi_{-, r, n}^{*} \, \diff \psi_{-, r, n} \, \e^{- \psi_{+, r, n}^{*}
    \psi_{+, r, n}} \\ \times \e^{- \psi_{-, r, n}^{*} \psi_{-, r, n}}
  \braket{\psi_{+, r, n} | \hat{D}_{r, n} | - \psi_{-, r, n}}
  \ket{\psi_{+, r, n}} \bra{- \psi_{-, r, n}},
\end{multline}
where $\psi_{\pm, r, n}$ and $\psi^{*}_{\pm, r, n}$ are independent Grassmann
variables,
\begin{equation}
  \diff \psi_{\pm, r, n} = \prod_{l = 1}^L \diff \psi_{\pm, r, l, n}, \quad
  \ket{\psi_{\pm, r, n}} = \e^{- \sum_{l = 1}^L \psi_{\pm, r, l, n}
    \hat{\psi}^{\dagger}_{\pm, r, l, n}} \ket{0},
\end{equation}
and as usual in the construction of fermionic Keldysh functional integrals we
flip the sign of Grassmann fields on the backward branch. To obtain a relation
between the coherent-state representations of the unnormalized density matrix at
times $t_n$ and $t_{n + 1}$, we insert
Eq.~\eqref{eq:coherent-state-representation} into Eq.~\eqref{eq:L-n}. We thus
have to evaluate matrix elements of the superoperator $\mathcal{L}_n$,
\begin{equation}
  \langle \psi_{+, r, n + 1} \rvert \mathcal{L}_n(\lvert \psi_{+, r, n} \rangle
  \langle - \psi_{-, r, n} \rvert) \lvert - \psi_{-, r, n + 1} \rangle.
\end{equation}
This is done using the fact that coherent ket and bra states are right and left
eigenstates of annihilation and creation operators, respectively. The final step
of the derivation is to exponentiate the second term in
Eq.~\eqref{eq:replica-linear-sme} in the coherent-state representation. Then,
using the discrete-time analog of the It\^o rule in Eq.~\eqref{eq:Ito-rules},
$\Delta W_{l, n} \Delta W_{l', n} = \delta_{l, l'} \Delta t$, and taking the
product over replicas as in Eq.~\eqref{eq:Z-R}, we obtain the $R$-replica
Keldysh partition function:
\begin{equation}
  \label{eq:Z-R-d-mu}
  Z_R = \int \diff \mu_{\Delta \mathbf{W}} \int \Diff[\psi^{*}, \psi] \,
  \e^{\imag \left( S_H[\psi^{*}, \psi] + \gamma S_M[\psi^{*}, \psi] \right)}.
\end{equation}
Writing sums over replicas and lattice sites as vector and matrix
multiplications, the Hamiltonian contribution to the action is
\begin{multline}
  \label{eq:S-H-discrete-time}
  S_H[\psi^{*}, \psi] \\ = \sum_{n = 0}^{N - 1} \left[ \Delta t \left( - \imag
      \frac{\psi_{+, n + 1}^{\dagger} - \psi_{+, n}^{\dagger}}{\Delta t}
      \psi_{+, n} - \imag \psi_{-, n}^{\dagger} \frac{\psi_{-, n + 1} - \psi_{-,
          n}}{\Delta t} \right) \right. \\ - i \left( \psi_{-, N}^{\dagger}
    \psi_{+, N} - \psi_{+,
      N}^{\dagger} \psi_{+, N} - \psi_{-, N}^{\dagger} \psi_{-, N} \right) \\
  \left. \vphantom{\sum_{n = 0}^{N - 1}} - \Delta t \left( \psi_{+, n +
        1}^{\dagger} H \psi_{+, n} - \psi_{-, n}^{\dagger} H \psi_{-, n + 1}
    \right) \right].
\end{multline}
In the continuous-time limit, the discrete differences in the second line can
formally be rewritten as derivatives, and the boundary terms in the third line
can be neglected. The elements of the matrix appearing in the last line are
given by
\begin{equation}
  \label{eq:Hamiltonian-matrix}
  H_{l, l'} = - J \left( \delta_{l + 1, l'} + \delta_{l, l' + 1} \right).
\end{equation}
Continuous measurements give rise to the following contribution to the Keldysh
action:
\begin{equation}
  \label{eq:S-M-Delta-W}
  \imag \gamma S_M[\psi^{*}, \psi] = - \sum_{l = 1}^L \sum_{n = 0}^{N - 1}
  \left[ \frac{\gamma \Delta t}{2} v^0_{l, n +1 , n} - \sqrt{\gamma}
    v^{\varphi}_{n + 1, n, l} \Delta W_{l, n} \right],
\end{equation}
where
\begin{equation}
  \label{eq:v-phi}
  v_{l, n, n'}^{\varphi} = \sum_{r = 1}^R \left( \psi^{*}_{+, r, l, n}
    \psi_{+, r, l, n'} \e^{\imag \varphi} + \psi^{*}_{-, r, l, n'} \psi_{-, r,
      l, n} \e^{- \imag \varphi} \right).  
\end{equation}

\subsection{Symmetry classification}
\label{sec:symmetry-classification}

The Keldysh partition function in Eq.~\eqref{eq:Z-R-d-mu} is similar to that of
free fermions in a disordered potential, with the increments $\Delta W_{l, n}$
playing the role of disorder~\cite{Kamenev2023, Altland2010a}. As such, the
Keldysh action of our model is amenable to the usual symmetry classification of
disordered systems~\cite{Altland1997, Evers2008}. Symmetries of the action have
to be taken into account in the derivation of the long-wavelength effective
field theory in terms of an NLSM~\cite{Jian2022, Fava2024, Poboiko2025,
  Starchl2025}.

We consider the continuous-time limit $t_n \to t$,
$\sum_{n = 0}^{N - 1} \Delta t \to \int_{t_0}^t \diff t$, and
$\Delta W_{l, n} \to \xi_l(t) \diff t$, which allows us to work with more
compact expressions. In this limit, the Keldysh action in
Eq.~\eqref{eq:Z-R-d-mu} can be written as
\begin{equation}
  S_H[\psi^{*}, \psi] + \gamma S_M[\psi^{*}, \psi] = \int_{t_0}^t \diff t' \,
  L[\psi^{*}(t'), \psi(t'), t'],
\end{equation}
with the Lagrangian
\begin{equation}
  L[\psi^{*}, \psi, t] = \psi^{\dagger} \left[ \left( \imag \partial_t
      - H \right) \sigma_z + \frac{\imag \gamma}{2} - \imag \sqrt{\gamma} \Xi(t)
  \e^{\imag \varphi \sigma_z} \right] \psi,
\end{equation}
where $\sigma_z$ is a Pauli matrix and the matrix
$\Xi_{l, l'}(t) = \xi_l(t) \delta_{l, l'}$ is diagonal in Keldysh, replica, and
real spaces. Following Ref.~\cite{Poboiko2025}, we introduce a chiral basis:
\begin{equation}
  \chi =
  \begin{pmatrix}
    \psi_+ \\ \psi_-
  \end{pmatrix},
  \qquad \chi^{\dagger} = \left( - \psi_-^{\dagger}, \psi_+^{\dagger} \right).
\end{equation}
In this basis, the Lagrangian takes the form
$L[\chi^{*}, \chi, t] = \chi^{\dagger} A(t) \chi$, where the Lagrangian matrix
$A(t)$ is given by
\begin{equation}
  \label{eq:Lagrangian-matrix}
  A(t) =
  \begin{pmatrix}
    0 & \imag \partial_t - H - \imag M^{*}(t) \\
    \imag \partial_t - H + \imag M(t) & 0
  \end{pmatrix},
\end{equation}
with the measurement matrix
\begin{equation}
  \label{eq:measurement-matrix}
  M(t) = \gamma/2 - \sqrt{\gamma} \Xi(t) \e^{\imag \varphi}.
\end{equation}
The Lagrangian matrix $A(t)$ has chiral symmetry,
$\left\{ A(t), \sigma_z \right\} = 0$. As discussed by Poboiko et
al.~\cite{Poboiko2025}, the Lagrangian matrix additionally has particle-hole
symmetry (PHS), $A(t) = - A^{\transpose}(t)$, if the following conditions are
satisfied: (i)~the hopping Hamiltonian $H$ has PHS, meaning that there is a
basis in which $H$ is skew-symmetric, $H = - H^{\transpose}$; and (ii)~the
measurement matrices $M(t)$ are real in the same basis, $M(t) = M^{*}(t)$. Let
us check whether our model obeys these conditions. The Hamiltonian
matrix~\eqref{eq:Hamiltonian-matrix} is rendered skew-symmetric by a local gauge
transformation, $H_{l, l'} \mapsto H_{l, l'} \imag^{l - l'}$.  This
transformation leaves the measurement matrix Eq.~\eqref{eq:measurement-matrix},
which is diagonal in real space, invariant. Therefore, the Lagrangian
matrix~\eqref{eq:Lagrangian-matrix} has PHS if the measurement
matrix~\eqref{eq:measurement-matrix} is real, which is the case only for
$\varphi = 0$. We conclude that for $\varphi = 0$, our model belongs to the
chiral orthogonal class BDI; instead, for $\varphi \neq 0$, it belongs to the
chiral unitary class AIII. This distinction affects the numerical value of the
prefactor of the weak-localization correction; however, it does not modify
entanglement properties and correlations qualitatively~\cite{Poboiko2025}.

\subsection{Average over trajectories and temporal regularization}
\label{sec:aver-over-traj}

We proceed with the analysis of the Keldysh partition
function~\eqref{eq:Z-R-d-mu} in the discrete-time formulation and take the
average over stochastic increments with measure
$\diff \mu_{\Delta \mathbf{W}} = \diff \Delta \mathbf{W} \, Q_{\Delta
  \mathbf{W}}$.
According to Eqs.~\eqref{eq:Q-Delta-W} and~\eqref{eq:S-M-Delta-W}, this average
amounts to preforming a Gaussian integral. We obtain
\begin{equation}  
  \e^{\imag \gamma \bar{S}_M[\psi^{*}, \psi]} = \int \diff \mu_{\Delta \mathbf{W}} \, \e^{\imag \gamma S_M[\psi^{*}, \psi]},
\end{equation}
where the averaged measurement action is given by
\begin{equation}
  \label{eq:S-M}
  \imag \bar{S}_M[\psi^{*}, \psi] = \frac{\Delta t}{2} \sum_{l = 1}^L
  \sum_{n = 0}^{N - 1} \left[ \left( v_{l, n + 1, n}^{\varphi} \right)^2 -
    v_{l, n + 1, n}^0 \right],
\end{equation}
with $v_{l, n, n'}^{\varphi}$ defined in Eq.~\eqref{eq:v-phi}.

The averaged measurement action~\eqref{eq:S-M} contains quartic vertices. We
wish to decouple these vertices by introducing fermionic bilinears as new
variables through a generalized Hubbard-Stratonovich
transformation. Furthermore, as detailed in Refs.~\cite{Poboiko2023,
  Starchl2025}, in the continuous-time limit, these bilinears of fermionic
fields should describe the symmetrized limit in which both fermionic fields are
evaluated at the same time $t$. Therefore, before taking the continuous-time
limit, the measurement action must be recast into a time-local form using the
``principal-value'' regularization~\cite{Poboiko2023, Starchl2025}.

As a first step, we rewrite the measurement action in a form that is strictly
local in time. This can be done by using properties of Grassmann integration,
which allow us to make the following replacements~\cite{Yang2023}:
\begin{equation}
  \begin{split}
    \psi^{*}_{+, r, l, n + 1} \psi_{+, r, l, n} & \to \psi^{*}_{+, r, l, n}
    \psi_{+, r, l, n} + 1, \\ \psi^{*}_{-, r, l, n} \psi_{-, r, l, n + 1} &
    \to \psi^{*}_{-, r, l, n} \psi_{-, r, l, n} + 1.
  \end{split}
\end{equation}
Note, however, that these replacements have to be made after expanding all
replica-diagonal products of polynomials of fields. We thus obtain
\begin{multline}
  \imag \bar{S}_M[\psi^{*}, \psi] = \frac{\Delta t}{2} \sum_{l = 1}^L \sum_{n =
    0}^{N - 1} \left\{ \sum_{r = 1}^R \left( \psi^{*}_{+, r, l, n} \psi_{+, r,
        l, n} \right. \right. \\ \left. + \psi^{*}_{-, r, l, n} \psi_{-, r, l,
      n} + 2 \psi^{*}_{+, r, l, n} \psi_{+, r, l, n} \psi^{*}_{-, r, l, n}
    \psi_{-, r, l, n} \right) \\ + \sum_{r \neq r'} \left[ \psi^{*}_{+, r, l, n}
    \psi_{+, r, l, n} \e^{\imag \varphi} + \psi^{*}_{-, r, l, n} \psi_{-, r, l,
      n} \e^{- \imag \varphi} + 2 \cos(\varphi) \right] \\
  \left. \vphantom{\sum_{r = 1}^R} \times \left[ \psi^{*}_{+, r', l, n} \psi_{+,
        r', l, n} \e^{\imag \varphi} + \psi^{*}_{-, r', l, n} \psi_{-, r', l, n}
      \e^{- \imag \varphi} + 2 \cos(\varphi) \right] \right\}. \\
  \tag*{\refstepcounter{equation}(\theequation)}
\end{multline}
In the replica limit $R \to 1$, only the first two lines yield a nonvanishing
contribution, which corresponds to the unconditional evolution of the average
state under dephasing as described by the master
equation~\eqref{eq:meq}~\cite{Yang2023}. Note that the unraveling phase
$\varphi$ appears only in the sum over $r$ and $r'$ with $r \neq r'$. This sum
vanishes in the replica limit, reflecting the fact that different unravelings
describe the same unconditional dynamics.

The second step toward the Hubbard-Stratonovich transformation is to implement
the principal-value regularization~\cite{Poboiko2023, Starchl2025}. To this end,
we first perform a Larkin-Ovchinnikov transformation. That is, we introduce new
fermionic fields that are defined by~\cite{Altland2010a, Kamenev2023}
\begin{equation}
  \label{eq:Larkin-Ovchinnikov-transformation}
  \psi_{1,2} = \frac{1}{\sqrt{2}} \left(\psi_+ \pm \psi_-\right), \qquad
  \psi^{*}_{1,2} = \frac{1}{\sqrt{2}} \left(\psi^{*}_+ \mp \psi^{*}_-\right).
\end{equation}
In the action expressed in terms of these fields, the principal-value
regularization amounts to adding to each vertex the sum of its partial Wick
contractions, with contractions of pairs of fields given by
\begin{equation}
  \begin{pmatrix}
    \langle \psi_{1, r, l, n} \psi^{*}_{1, r', l', n'} \rangle &
    \langle \psi_{1, r, l, n} \psi^{*}_{2, r', l', n'} \rangle \\
    \langle \psi_{2, r, l, n} \psi^{*}_{1, r', l', n'} \rangle &
    \langle \psi_{2, r, l, n} \psi^{*}_{2, r', l', n'} \rangle
  \end{pmatrix}
  = \frac{1}{2} \delta_{r, r'} \delta_{l, l'} \delta_{n, n'} \sigma_x.
\end{equation}
After this step, we take the continuous-time limit. Using the shorthand notation
$\int \diff^2 \mathbf{x} = \sum_{l=1}^L \int_{t_0}^t \diff t'$, the averaged
measurement action can be written as
\begin{equation}
  \bar{S}_M[\psi^{*}, \psi] = \int \diff^2 \mathbf{x} \, \mathcal{L}_M[\psi^{*}_l(t'), \psi_l(t')].
\end{equation}
The measurement Lagrangian, expressed in terms of
$\psi_r = \left( \psi_{1, r}, \psi_{2, r} \right)^{\transpose}$, is given by
\begin{multline}
  \label{eq:L-M-psi}
  \imag \mathcal{L}_M[\psi^{*}, \psi] = \frac{R}{2} \left[ \left( R - 1 \right)
    \cos(\varphi)^2 - \frac{1}{2} \right] \\ + \left( R - 1 \right)
  \cos(\varphi) \sum_{r = 1}^R \psi_r^{\dagger} X_{\varphi} \psi_r + \frac{1}{2}
  \left( \sum_{r = 1}^R \psi_r^{\dagger} X_{\varphi} \psi_r \right)^2,
\end{multline}
where $X_{\varphi}$ is a matrix in Keldysh space,
\begin{equation}
  \label{eq:X-phi}
  X_{\varphi} = \cos(\varphi) \sigma_x + \imag \sin(\varphi) = \sigma_x
  \e^{\imag \varphi \sigma_x}.
\end{equation}
In this notation, the Hamiltonian action~\eqref{eq:S-H-discrete-time} becomes
\begin{equation}
  \label{eq:S-H-continuous-time}
  S_H[\psi^{*}, \psi] = \sum_{r=1}^R \int_{t_0}^t \diff t' \psi_r^{\dagger}(t')
  G_0^{-1} \psi_r(t'),
  \end{equation}
with the free fermion Green's function
\begin{equation}
  G_0^{-1} = \imag \partial_t - H.
\end{equation}

\subsection{Generalized Hubbard-Stratonovich transformation}

We are now in a position to decouple the quartic vertex in the measurement
Lagrangian~\eqref{eq:L-M-psi}. The vertex is local in both space and time, and
can thus be decoupled in terms of local fermionic bilinears, which we collect in
the matrix
$\mathcal{G}_l(t) = - \imag \psi_l(t) \psi^{*}_l(t)$~\cite{Poboiko2023,
  Starchl2025}. We note that since we have imposed the principal-value
regularization, the expectation value
$\left\langle \mathcal{G}(t) \right\rangle$ is equal to the fermionic Green's
function in the symmetrized limit of equal time arguments of the fields
$\psi(t)$ and $\psi^{*}(t)$,
\begin{equation}
  \label{eq:G-symmetric-limit}
  \left\langle \mathcal{G}_l(t) \right\rangle = - \frac{\imag}{2} \left(
    \langle \psi_l(t) \psi^{*}_l(t + 0^+) \rangle + \langle \psi_l(t) \psi^{*}_l(t - 0^+) \rangle \right).
\end{equation}
To perform the Hubbard-Stratonovich transformation of the Keldysh partition
function,
\begin{equation}
  \label{eq:Z-R-d-psi}
  Z_R = \int \Diff[\psi^{*}, \psi] \, \e^{\imag \left( S_H[\psi^{*}, \psi] +
      \gamma \bar{S}_M[\psi^{*}, \psi] \right)},
\end{equation}
we insert the identity
\begin{equation}
  1 = \int \Diff[\mathcal{G}, \Sigma] \, \e^{- \imag \Tr \left( \mathcal{G}
      \Sigma \right) - \psi^{\dagger} \Sigma \psi},
\end{equation}
where $\mathcal{G}$ and $\Sigma$ are Hermitian $2R \times 2R$ matrix fields. The
trace $\Tr$ acts in Keldysh, replica, lattice, and time spaces. Integration over
$\Sigma$ yields a delta functional that enforces the identity
$\mathcal{G} = - \imag \psi \psi^{\dagger}$, which allows us to rewrite the quartic
vertex in Eq.~\eqref{eq:L-M-psi} in terms of $\mathcal{G}$. In principle, this
can be done in several ways which correspond to decoupling the vertex in
different channels. However, as discussed by Poboiko et al.~\cite{Poboiko2023},
for $\mathcal{G}$ to describe long-wavelength fluctuations, the vertex must be
decoupled simultaneously in all possible channels. This is achieved by taking
the sum over all Wick contractions of the measurement Lagrangian, which yields
\begin{multline}
  \label{eq:L-M-G}
  \imag \mathcal{L}_M[\mathcal{G}] = \frac{R}{2} \left[ \left( R - 1 \right)
    \cos(\varphi)^2 - \frac{1}{2} \right] \\ - \imag \left( R - 1 \right)
  \cos(\varphi) \tr \! \left( X_{\varphi} \mathcal{G} \right) + \frac{1}{2}
  \left\{ \tr \! \left[ \left( X_{\varphi} \mathcal{G} \right)^2 \right] - \tr
    \! \left( X_{\varphi} \mathcal{G} \right)^2 \right\}.
\end{multline}
As we have noted earlier, in the replica limit $R \to 1$, corresponding to the
unconditional dynamics described by the master equation~\eqref{eq:meq}, the
unraveling phase $\varphi$ has to drop out of the measurement
Lagrangian. Indeed, in this case, $\mathcal{G}$ reduces to a $2 \times 2$ matrix
and we find
\begin{equation}
  \label{eq:L-M-G-R=1}
  \imag \mathcal{L}_M[\mathcal{G}] = - \frac{1}{4} + \detK \! \left(
    \mathcal{G} \right).
\end{equation}
Interestingly, this result coincides with the one for random projective
measurements of occupation numbers, given in Eq.~(23) of
Ref.~\cite{Poboiko2023}, up to a factor of $1/2$.

Now that we have expressed the measurement Lagrangian in terms of $\mathcal{G}$,
the remaining integral over fermionic fields is Gaussian and can be performed
straightforwardly. We obtain
\begin{equation}
  \label{eq:Z-R-d-G-d-Sigma}
  Z_R = \int \Diff[\mathcal{G}, \Sigma] \, \e^{\imag S[\mathcal{G}, \Sigma]},
\end{equation}
where the action is now given by
\begin{equation}
  \label{eq:S-G-Sigma}
  S[\mathcal{G}, \Sigma] = S_0[\mathcal{G}, \Sigma] + \gamma \int \diff^2
  \mathbf{x} \, \mathcal{L}_M[\mathcal{G}_l(t')],
\end{equation}
with
\begin{equation}
  \label{eq:S-0-G-Sigma}
  \imag S_0[\mathcal{G}, \Sigma] = \Tr \! \left[ \ln\!\left(\imag \partial_t - H +
      \imag \Sigma \right) -\imag \mathcal{G} \Sigma \right].
\end{equation}

\section{Gaussian theory}
\label{sec:gaussian-theory}

Our further analysis of the replica field theory proceeds in two steps: First,
we study Gaussian fluctuations around a particular saddle point of the
action~\eqref{eq:S-G-Sigma}; second, in Sec.~\ref{sec:NLSM} below, we extend
this analysis to nonlinear fluctuations. The Gaussian approximation is valid for
$\gamma \ll J$ and on short to intermediate length scales~\cite{Poboiko2023}. On
large scales, strong fluctuations of Goldstone modes, which are described by an
NLSM, lead to deviations from the Gaussian theory.

\subsection{Saddle points and fluctuation expansion}

In the Gaussian approximation, we expand the Keldysh action~\eqref{eq:S-G-Sigma}
to second order in fluctuations around a particular saddle point that describes
the stationary state, meaning that $\mathcal{G}$ and $\Sigma$ are spatially
homogeneous and time-independent. To find such a saddle point, we first take the
variational derivative of the Keldysh action with respect to $\Sigma$, which
yields
\begin{equation}
  \mathcal{G} = \int_{-\pi}^{\pi} \frac{\diff q}{2 \pi} \int_{-\infty}^{\infty}
  \frac{\diff \omega}{2 \pi} \frac{1}{\omega - \xi_q + \imag \Sigma},
\end{equation}
where $\xi_q = - 2 J \cos(q)$. As it stands, the integral over frequencies is
divergent. A regularization that is consistent with the symmetric limit of equal
times in Eq.~\eqref{eq:G-symmetric-limit} is obtained by inserting the factor
$\bigl( \e^{\imag \omega t} + \e^{- \imag \omega t} \bigr)\big/2$ and taking the
limit $t \to 0^+$ after integration. The integral over the matrix-valued
integrand can then be evaluated by writing
$\Sigma = \mathcal{V} \lambda \mathcal{V}^{-1}$ where $\lambda$ is a diagonal
matrix. This leads to
\begin{equation}
  \label{eq:G-Q}
  \mathcal{G} = - \imag Q/2, \quad Q = \sgn \! \left[ \Re(\Sigma) \right] =
  \mathcal{V} \sgn \! \left[\Re(\lambda)\right] \mathcal{V}^{-1}.
\end{equation}
Next, we take the variational derivative of the Keldysh action with respect to
$\mathcal{G}$. Setting $\mathcal{G} = - \imag Q/2$, we obtain
\begin{equation}
  \label{eq:Sigma-Q}
  \Sigma = - \gamma \left\{ \left( R - 1 \right) \cos(\varphi) + \frac{1}{2}
    \left[ X_{\varphi} Q - \tr \! \left( X_{\varphi} Q \right) \right] \right\}
  X_{\varphi}.
\end{equation}
Any matrix $Q$ that obeys the condition $Q^2 = 1$ yields a solution of
Eqs.~\eqref{eq:G-Q} and~\eqref{eq:Sigma-Q}. The form of $Q$ can be restricted
further on physical grounds: We expect $Q$ to be replica-symmetric,
$Q = Q_K \otimes 1_R$; moreover, we expect $Q_K$ to obey the usual causality
structure of Green's functions in Keldysh field theory~\cite{Kamenev2023}. For a
given density $n = N/L$, such a saddle point is given by $Q_K = \Lambda$ with
\begin{equation}
  \label{eq:lambda}
  \Lambda =
  \begin{pmatrix}
    1 & 2 \left( 1 - 2 n \right) \\ 0 & - 1
  \end{pmatrix}.
\end{equation}
To see that this saddle point yields the correct density of fermions, note that
from Eq.~\eqref{eq:G-symmetric-limit} and the relation between $\mathcal{G}$ and
$Q$ in Eq.~\eqref{eq:G-Q} it follows that the ``classical'' Keldysh component of
the density for a single replica is given by
\begin{equation}
  \label{eq:rho-classical}
  \rho_r = \frac{1}{4} \trK \! \left( 1 - \sigma_x Q_{r, r} \right).
\end{equation}
Therefore, at the replica-symmetric saddle point $Q = \Lambda \otimes 1_R$, the
density is indeed $\rho_r = n$. Furthermore, note that $\Lambda = F \sigma_z F$
with
\begin{equation}
  \label{eq:F}
  F = F^{-1} =
  \begin{pmatrix}
    1 & 1 - 2 n \\ 0 & - 1
  \end{pmatrix}.
\end{equation}
As such, $Q$ takes the form given in Eq.~\eqref{eq:G-Q}, with
$\mathcal{V} = F \otimes 1_R$ and $\lambda = \sigma_z \otimes 1_R$. Finally, we
note that Eq.~\eqref{eq:Sigma-Q} simplifies further for a replica-symmetric
saddle point, as can be seen by using the definition of the matrix $X_{\varphi}$
given in Eq.~\eqref{eq:X-phi}. Upon setting $R = 1$ in the numerical prefactor
of the first term in Eq.~\eqref{eq:Sigma-Q}, we find that Eqs.~\eqref{eq:G-Q}
and~\eqref{eq:Sigma-Q} can be written as
\begin{equation}
  \label{eq:G-Sigma-Q-Lambda}
  \mathcal{G} = - \imag Q/2, \qquad \Sigma = \gamma Q/2, \qquad Q = \Lambda.
\end{equation}
Here and in the following, we omit the identity in replica space for the sake of
brevity. The fermionic Green's functions corresponding to the saddle
point~\eqref{eq:G-Sigma-Q-Lambda} are~\cite{Poboiko2023}
\begin{equation}
  \label{eq:Green's-functions}
  \begin{split}
    G^{R/A}_q(\omega) & = \frac{1}{\omega - \xi_q \pm \imag \gamma/2}, \\
    G^K_q(\omega) & = \left( 1 - 2 n \right) \left[ G^R_q(\omega) -
      G^A_q(\omega) \right].
  \end{split}
\end{equation}
These Green's functions describe fermions at infinite temperature. Gaussian
fluctuations around the saddle point~\eqref{eq:G-Sigma-Q-Lambda} can be
parametrized as
\begin{equation}
  \label{eq:G-Sigma-fluctuations}
  \mathcal{G} = -\imag \left( \Lambda + \dQG{} \right) \! \big/2, \qquad \Sigma
  = \gamma \left( \Lambda + \dQS{} \right) \! \big/2,
\end{equation} 
with Hermitian $2 R \times 2 R$ matrices $\dQG$ and $\dQS$. The Gaussian theory
is obtained by expanding the Keldysh action~\eqref{eq:S-G-Sigma} to second
order in $\dQG$ and $\dQS$ as detailed in Refs.~\cite{Poboiko2023,
  Starchl2025}.

\subsection{Connected density correlation function}

As explained in Sec.~\ref{sec:line-stoch-mast}, nontrivial effects of
measurements manifest only in nonlinear observables. Here we focus on the
equal-time connected density correlation function. Its definition in the
operator formalism and the corresponding expression in terms of field
expectation values are given by
\begin{equation}
  \label{eq:C-l-l-prime-t}
  \begin{split}
    C_{l, l'}(t) & = \frac{1}{2} \overline{\left\langle \left\{ \hat{n}_l(t),
          \hat{n}_{l'}(t) \right\} \right\rangle} - \overline{\left\langle
        \hat{n}_l(t) \right\rangle \left\langle \hat{n}_{l'}(t) \right\rangle} \\
    & = C_{r, r, l, l'}(t, t) - C_{r, r', l, l'}(t, t),
  \end{split}
\end{equation}
where the correlation function of density fluctuations is
\begin{equation}
  \label{eq:C-r-r-prime}
  C_{r, r', l, l'}(t, t') = \langle \delta \rho_{r, l}(t) \delta \rho_{r',
    l'}(t') \rangle.
\end{equation}
Due to the symmetry of the formalism under permutations of replicas, the
specific choices of $r$ and $r'$ in Eq.~\eqref{eq:C-l-l-prime-t} are irrelevant
as long as $r \neq r'$. According to Eq.~\eqref{eq:rho-classical}, density
fluctuations are related to $\dQG$ by
\begin{equation}
  \delta \rho_{r, l}(t) = - \frac{1}{4} \trK \! \left[ \sigma_x \delta
    Q_{\mathcal{G}, r, r, l}(t) \right].
\end{equation}
The correlation function~\eqref{eq:C-l-l-prime-t} can be obtained from the
Keldysh partition function~\eqref{eq:Z-R-d-G-d-Sigma} by introducing sources
that couple to density fluctuations~\cite{Poboiko2023} and performing the
Gaussian integral over $\delta Q_{\mathcal{G}}$ and $\delta Q_{\Sigma}$.

The main technical difficulty in calculating the density correlation
function~\eqref{eq:C-l-l-prime-t} is accounting for the boundary condition
imposed by stopping the measurement process at a finite time $t$, which requires
us to treat matrices in time space as noncommuting~\cite{Poboiko2023}. Without
loss of generality, we set the final time to $t = 0$. Then, we define the
product of two matrices in time space as
\begin{equation}
  \label{eq:time-space-matrix-multiplication}
  \left( A B \right)(t, t') = \int_{-\infty}^0 \diff t'' \, A(t, t'') B(t'', t'),
\end{equation}
where we leave Keldysh, replica, and lattice-site indices of $A$ and $B$
unspecified. Using this convention, we obtain the density correlation
function~\eqref{eq:C-l-l-prime-t} in the form
\begin{equation}
  \label{eq:C-l-Gaussian-full-numerical}
  C_{l, l'}(t) = \left\{ \mathcal{C} \left[ 1 + 2 \gamma \cos(2 \varphi)
      \mathcal{C} \right]^{-1} \right\}_{l, l'} \!\! (t, t).
\end{equation}
For simplicity, we focus here on a half-filled chain, $n = 1/2$. We do not
expect the physics to depend qualitatively on the value of $n$, apart from the
limiting cases $n = 0$ and $n = 1$ in which the model becomes trivial. In
Eq.~\eqref{eq:C-l-Gaussian-full-numerical}, $\mathcal{C}$ is the density
correlation function found previously in Refs.~\cite{Poboiko2023, Starchl2025}
for different types of measurement processes, with and without a conserved
number of particles, but with a fixed mean density of $n = 1/2$. It is given by
\begin{equation}
  \label{eq:C-pi/4-case}
  \mathcal{C} = \frac{1}{2} \left( \mathcal{B}_+ + \gamma \mathcal{B}_- L
    \mathcal{B}_- \right).
\end{equation}
The quantities $\mathcal{B}_{\pm}$ and $L$ are most conveniently specified in
momentum space:
\begin{equation}
  \mathcal{B}_{\pm, q}(t) = \frac{1}{2} \left[ \mathcal{B}_q(t) \pm
    \mathcal{B}_q(- t) \right],
\end{equation}
where
\begin{equation}
  \mathcal{B}_q(t) = \theta(t) \e^{- \gamma t} J_0 \! \left[ 4 J t \sin(q/2)
  \right],
\end{equation}
which is related to the Green's functions~\eqref{eq:Green's-functions} via
$\mathcal{B}_l(t) = G^R_l(t) G^A_{-l}(- t)$, and where $J_0$ is a Bessel
function of the first kind~\cite{Poboiko2023}; furthermore, $L_q(t, t')$ is the
solution of a Wiener-Hopf integral equation,
\begin{equation}
  \label{eq:Wiener-Hopf}
  L_q(t, t') - \gamma \int_{-\infty}^0 \diff t'' \, \mathcal{B}_{+,
    q}(t - t'') L_q(t'', t') = \delta(t - t').
\end{equation}
Through rescaling time, the integral equation~\eqref{eq:Wiener-Hopf} can be seen
to depend on a single parameter:
\begin{equation}
  \label{eq:u-l-0}
  u = 2 l_0 \sin(q/2) \sim q l_0, \qquad l_0 = 2 J/\gamma.
\end{equation}
We have solved Eqs.~\eqref{eq:C-l-Gaussian-full-numerical},
\eqref{eq:C-pi/4-case}, and~\eqref{eq:Wiener-Hopf} numerically by discretizing
time $t$ such that integrals over time as in
Eq.~\eqref{eq:time-space-matrix-multiplication} are converted into ordinary
matrix multiplications. Details of the numerical procedure are provided in
Appendix~\ref{sec:wienerhopf}.

Interestingly, the leading asymptotic behavior of the momentum-space transform
$C_q(t)$ of Eq.~\eqref{eq:C-l-Gaussian-full-numerical} for both small and large
momenta is correctly reproduced in the bulk approximation, in which the range of
integration in Eq.~\eqref{eq:time-space-matrix-multiplication} is extended to
$t \in (-\infty, \infty)$---that is, the boundary condition in time is
ignored---and the parameter $u$~\eqref{eq:u-l-0} is replaced by twice its
value~\cite{Poboiko2023}. We find
\begin{equation}
  \label{eq:C-q-Gaussian-bulk}
  \frac{C_q}{g_{\varphi, 0} q} = \frac{\tilde{c}(u)}{u},
\end{equation}
where we define
\begin{equation}
  \label{eq:g-phi-0}
  g_{\varphi, 0} = \frac{l_0 n \left( 1 - n \right)}{\cos(\varphi)},
\end{equation}
and the function $\tilde{c}(u)$ is given by
\begin{multline}
  \label{eq:c-tilde}
  \tilde{c}(u) = \frac{2 \cos(\varphi)}{\pi} \int_0^{\infty} \diff v \left\{
    \Re[b(2 u, v)] - \abs{b(2 u, v)}^2 \right\} \\ \times \left\{ 1 + f^2
    \abs{b(2 u, v)}^2 -
    \left( 1 + f^2 \right) \Re[b(2 u, v)] \right. \\
  \left. \vphantom{\left( 1 + f^2 \right)} + 4 n \left( 1 - n \right) \cos(2
    \varphi) \left[ \Re[b(2 u, v)] - \abs{b(2 u, v)}^2 \right] \right\}^{-1},
\end{multline}
with $f = 1 - 2n$ and
\begin{equation}
  b(u, v) = \left[ \left( 1 - \imag v \right)^2 + u^2 \right]^{-1/2}.
\end{equation}
Here and in the following, we omit the time argument of $C_q$, which becomes
time-independent in the steady state. The leading asymptotic behavior of
$\tilde{c}(u)$ for $n = 1/2$ is 
\begin{equation}
  \label{eq:c-tilde-asymptotic}
  \tilde{c}(u) \sim
  \begin{cases}
    u, & u \to 0, \\
    \cos(\varphi), & u \to \infty.
  \end{cases}
\end{equation}
We compare the full numerical solution for the density correlation
function~\eqref{eq:C-l-Gaussian-full-numerical} with the bulk
approximation~\eqref{eq:C-q-Gaussian-bulk} in Sec.~\ref{sec:numerics-vs-theroy}
below. The discrepancies for intermediate values of $u$ indicate that higher
orders in asymptotic expansions of the bulk
solution~\eqref{eq:c-tilde-asymptotic} do not agree with the full solution. As a
result, the bulk approximation lacks the accuracy needed to determine the
weak-localization correction, even though the overall agreement between the bulk
approximation and the full numerical solution is good.

Transforming the momentum-space correlation
function~\eqref{eq:C-q-Gaussian-bulk} back to real space, the leading asymptotic
behavior in Eq.~\eqref{eq:c-tilde-asymptotic} for $u \to 0$ implies that $C_l$
decays algebraically on large scales as expected for a critical
phase~\cite{Alberton2021, Eissler2025}:
\begin{equation}
  \label{eq:C-l-Gaussian-bulk-approximation}
  \frac{l_0^2 C_l}{g_{\varphi, 0}} \sim - \frac{1}{\pi y^2}, \qquad y =
  \frac{l}{l_0} \to \infty.
\end{equation}
Recall that the Green's functions~\eqref{eq:Green's-functions} associated with
the replica-symmetric saddle point~\eqref{eq:G-Sigma-Q-Lambda} describe a Fermi
gas at infinite temperature, with vanishing
correlations~\eqref{eq:C-l-l-prime-t} for $l \neq l'$. The result in
Eq.~\eqref{eq:C-l-Gaussian-bulk-approximation} shows that including Gaussian
fluctuations around the saddle point brings about a first qualitative
modification of correlations on scales $l \gtrsim l_0$. A further modification
on large scales results from nonlinear fluctuations of Goldstone modes as we
discuss in Sec.~\ref{sec:NLSM}.

\subsection{Entanglement entropy}
\label{sec:entanglement-entropy-Gaussian}

Having discussed the connected density correlation
function~\eqref{eq:C-l-l-prime-t}, we now turn to another important example of a
nonlinear observable: the entanglement entropy. Note that in quantum
trajectories, the system is always in a pure state
$\hat{\rho} = \ket{\psi} \bra{\psi}$. The reduced density matrix of a subsystem
consisting of $\ell$ contiguous lattice sites is
$\hat{\rho}_{\ell} = \tr_{L - \ell}(\hat{\rho})$, where the trace is taken over
the remaining $L - \ell$ sites, and the von Neumann entanglement entropy of the
subsystem is defined as
\begin{equation}
  \label{eq:entanglement-entropy}
  \svn{\ell} = - \overline{\tr \! \left[ \hat{\rho}_{\ell}
      \ln\!\left(\hat{\rho}_{\ell}\right) \right]}.
\end{equation}
To calculate the entanglement entropy, we use another property of the quantum
trajectories described by Eq.~\eqref{eq:sse}: Since the generator of the
stochastic dynamics is quadratic in fermionic operators, an initially Gaussian
state remains Gaussian at all times; and for Gaussian states, the entanglement
entropy is determined by the fluctuations of the number of particles in the
subsystem, $\hat{N}_{\ell} = \sum_{l = 1}^{\ell} \hat{n}_l$, through an
expansion in cumulants~\cite{Klich2009, Song2011, Song2012, Thomas2015,
  Burmistrov2017}:
\begin{equation}
  \label{eq:entropy-cumulant}
  \svn{\ell} = 2 \sum_{k = 1}^{\infty}  \zeta(2k) C_{\ell}^{(2 k)} = \frac{\pi^2}{3}
  C_{\ell}^{(2)} +  \frac{\pi^4}{45} C_{\ell}^{(4)} + \dotsb.
\end{equation}
The coefficients in this expansion are given in terms of the Riemann zeta
function $\zeta(k)$. As demonstrated below, the numerical data for $\svn{\ell}$
is well approximated by keeping only the contribution from the second cumulant,
which is given by
\begin{equation}
  \label{eq:cumulant}
  C_{\ell}^{(2)} = \overline{\langle ( \hat{N}_{\ell} - \langle \hat{N}_{\ell}
    \rangle )^2 \rangle}.
\end{equation}
The second cumulant can be expressed in terms of the conditional density
correlation function~\eqref{eq:C-l-l-prime-t}, which leads to
\begin{equation}
  \label{eq:entanglement-entropy-correlation-function}
  \svn{\ell} = \frac{\pi^2}{3} \sum_{l, l' =
    1}^{\ell} C_{l - l'} = \frac{2 \pi}{3} \int_0^{\infty} \frac{\diff q}{q^2}
  \, C_q \left[ 1 - \cos(q \ell) \right].
\end{equation}
Inserting here the Gaussian correlation function in the bulk
approximation~\eqref{eq:C-q-Gaussian-bulk} and using the asymptotic behavior for
$n = 1/2$ given in Eq.~\eqref{eq:c-tilde-asymptotic}, we find the behavior of
$\svn{\ell}$ for small and large values of the subsystem size $\ell$:
\begin{equation}
  \label{eq:S-l-asymptotic}
  S_{\ell} \sim
  \begin{cases}
    \pi \cos(\varphi) \ell/(2 l_0), & \ell \to 0, \\
    (2 \pi g_{\varphi, 0}/3) \ln(\ell/l_0), & \ell \to \infty.
  \end{cases}
\end{equation}
On short scales $\ell \lesssim l_0$, we expect the system to behave as a Fermi
gas at infinite temperature. In particular, we expect the entanglement entropy
to obey a volume law, $S_{\ell} \sim \ln(2) \ell$ for $n = 1/2$. This
expectation is not borne out by the Gaussian result~\eqref{eq:S-l-asymptotic},
which indicates that on short scales contributions from higher cumulants in
Eq.~\eqref{eq:entropy-cumulant} cannot be discarded~\cite{Poboiko2023}.

On large scales $\ell \gg l_0$, the Gaussian theory predicts logarithmic growth
of the entanglement entropy as characteristic for a 1D conformal field theory
(CFT)~\cite{DiFrancesco1997}. To characterize deviations from CFT behavior which
occur on short scales and, as we below, also on large scales due to nonlinear
fluctuations, we define the scale-dependent effective central
charge~\cite{Starchl2025}:
\begin{equation}
  \label{eq:effective-central-charge}
  c_{\ell} = 3 \frac{\partial \svn{\ell}}{\partial \! \ln(\ell)}.
\end{equation}
According to the asymptotic behavior of the Gaussian entanglement entropy in
Eq.~\eqref{eq:S-l-asymptotic}, on short scales, $c_{\ell} \sim \ell$ vanishes
for $\ell \to 0$; on large scales, we find
\begin{equation}
  \label{eq:effective-central-charge-Gaussian}
  c_{\ell} \sim 2 \pi g_{\varphi, 0}, \qquad \ell \to \infty,
\end{equation}
with $g_{\varphi, 0}$ given in Eq.~\eqref{eq:g-phi-0}. As anticipated at the
beginning of Sec.~\ref{sec:gaussian-theory}, we next extend our analysis to
include nonlinear fluctuations, described by an NLSM. These fluctuations induce
a renormalization group (RG) flow of $g_{\varphi, 0}$ that causes the effective
central charge to flow to zero on large scales, indicating that logarithmic
growth of the entanglement entropy crosses over to area-law scaling.

\section{Nonlinear sigma model}
\label{sec:NLSM}

According to the symmetry analysis of the Keldysh action in
Sec.~\ref{sec:symmetry-classification}, our model belongs to class BDI for
$\varphi = 0$ and to class AIII for $\varphi \neq 0$.
References~\cite{Poboiko2025, Starchl2025} present the derivation of the NLSM
for class BDI under different measurement schemes, while the derivation for
class AIII is detailed in Refs.~\cite{Poboiko2023, Starchl2025, Poboiko2024,
  Chahine2024}. Conceptually, the derivation is the same for both symmetry
classes. Here, we briefly summarize the key steps, focusing on $\varphi \neq 0$
where particle-hole symmetry is broken, and indicate modifications due to
particle-hole symmetry for $\varphi = 0$ where relevant.

\subsection{Target manifold}
\label{sec:target-manifold}

Due to the continuous symmetries of the action, the saddle
point~\eqref{eq:G-Sigma-Q-Lambda} is not isolated but rather embedded in a
manifold of saddle points. The NLSM describes long-wavelength fluctuations of
the Goldstone modes within this manifold. Consequently, to derive the NLSM, we
first have to determine the structure of its target manifold---that is, the
manifold of saddle points---and find an explicit parameterization.

To construct the NLSM target manifold, we use the replica-symmetric saddle
point~\eqref{eq:G-Sigma-Q-Lambda} as the base point to which we apply
transformations $\mathcal{R} \in \mathrm{U}(2 R)$. For the time being, we set
$n = 1/2$ such that Eq.~\eqref{eq:lambda} reduces to $\Lambda = \sigma_z$. We
thus obtain
\begin{equation}
  \label{eq:saddle-point-manifold-n=1/2}
  \mathcal{G} = - \imag Q/2, \qquad \Sigma = \gamma Q/2, \qquad Q = \mathcal{R}
  \sigma_z \mathcal{R}^{-1}.
\end{equation}
For this to parameterize the NLSM manifold, we require the transformations
$\mathcal{R}$ to obey two conditions: (i)~They should be symmetries of the
Keldysh action. This guarantees that $Q = \sigma_z$ being a saddle point entails
$Q = \mathcal{R} \sigma_z \mathcal{R}^{-1}$ is one as well. (ii)~They should
transform $\Lambda = \sigma_z$ nontrivially. Regarding condition (i), note that
any transformation $\mathcal{R}$ is a symmetry of the contribution to the
Keldysh action given in Eq.~\eqref{eq:S-0-G-Sigma}. Therefore, which
transformations $\mathcal{R}$ are symmetries is determined solely by the
measurement Lagrangian.  As can be seen from Eqs.~\eqref{eq:L-M-G}
and~\eqref{eq:L-M-G-R=1}, symmetries of the measurement Lagrangian are different
for $R = 1$ and $R > 1$.

We consider first the replica-symmetric case with $R = 1$. Condition (i) does
not restrict the NLSM manifold: Arbitrary transformations
$\mathcal{R}_0 \in \mathrm{U}(2)$ are symmetries of Eq.~\eqref{eq:L-M-G-R=1}. To
restrict the group $\mathrm{U}(2)$ to transformations that obey condition (ii),
we factor out the group of transformations $\mathcal{R}_0$ that leave the saddle
point $\Lambda = \sigma_z$ invariant. These transformations are determined by
\begin{equation}
  \left[ \mathcal{R}_0, \sigma_z \right] = 0 \quad \Longleftrightarrow
  \quad \mathcal{R}_0 =
  \begin{pmatrix}
    \e^{\imag \phi_1} & 0 \\ 0 & \e^{\imag \phi_2}
  \end{pmatrix}.
\end{equation}
Matrices $\mathcal{R}_0$ with this structure form the group
$\mathrm{U}(1) \times \mathrm{U}(1)$. The replica-symmetric NLSM manifold for
$n = 1/2$ can thus be parameterized as
\begin{equation}
  Q_0 = \mathcal{R}_0 \sigma_z \mathcal{R}_0^{-1},
\end{equation}
where
$\mathcal{R}_0 \in \mathrm{U}(2)/\mathrm{U}(1) \times \mathrm{U}(1) \simeq
\mathrm{S}^2.$ The generalization to arbitrary densities $n \neq 1/2$ reads
\begin{equation}
  \label{eq:Q-0}
  Q_0 = F \mathcal{R}_0 \sigma_z \mathcal{R}_0^{-1} F,
\end{equation}
with $F$ given in Eq.~\eqref{eq:F}. Note that when $Q_0$ is inserted in the
Keldysh action~\eqref{eq:S-G-Sigma}, $F$ drops out due to $F = F^{-1}$. However,
inserting $Q_0$ in the expression for the density in
Eq.~\eqref{eq:rho-classical} shows that Eq.~\eqref{eq:Q-0} indeed describes a
system with density $n$.

Having discussed the replica-symmetric manifold with $R = 1$, we now turn to the
replicon manifold with $R > 1$. The replica-symmetric manifold can be embedded
in the replicon manifold by setting
\begin{equation}
  \label{eq:G-Sigma-Q}
  \mathcal{G} = - \imag Q/2, \qquad \Sigma = \gamma Q/2, \qquad Q = \mathcal{R}
  Q_0 \mathcal{R}^{-1}.
\end{equation}
with $Q_0$ given in Eq.~\eqref{eq:Q-0}. Here, in addition to (i)~being a
symmetry and (ii)~transforming $Q_0$ nontrivially, the transformations
$\mathcal{R}$ should (iii)~reduce to the identity for $R = 1$. A necessary
condition for $\mathcal{R}$ to obey requirement (iii) is that
$\mathcal{R} \in \mathrm{SU}(2 R)$, since $\mathrm{U}(1)$ phase factors are
already included in the replica-symmetric manifold. To determine the form of the
transformations $\mathcal{R} \in \mathrm{SU}(2 R)$ that obey conditions (i) and
(ii), it is convenient to use the representation
$\mathcal{R} = M \mathcal{R}' M$ with
$M = M^{\dagger} = M^{-1} = (\sigma_x + \sigma_z)/ \sqrt{2}$. Note that the
matrix $M$ describes the Larkin-Ovchinnikov
transformation~\eqref{eq:Larkin-Ovchinnikov-transformation} up to a sign flip of
$\psi_-^{*}$. When we insert this representation of $\mathcal{R}$ into the
measurement Lagrangian Eq.~\eqref{eq:L-M-G}, the matrix $X_{\varphi}$ that
occurs in the Lagrangian is transformed into
\begin{equation}
  Z_{\varphi} = M X_{\varphi} M = \cos(\varphi) \sigma_z + \imag \sin(\varphi) =
  \sigma_z \e^{\imag \varphi \sigma_z}.
\end{equation}
We thus see that $\mathcal{R} = M \mathcal{R}' M$ is a symmetry of the action if
\begin{equation}
  \label{eq:R-prime-sigma-z}
  \left[ \mathcal{R}', Z_{\varphi} \right] = 0 \quad \Longleftrightarrow \quad \left[
    \mathcal{R}', \sigma_z \right] = 0.
\end{equation}
This condition implies that $\mathcal{R}'$ must be block-diagonal in Keldysh
space,
\begin{equation}
  \label{eq:R-+-R--}
  \mathcal{R}' =
  \begin{pmatrix}
    \mathcal{R}_+' & 0 \\ 0 & \mathcal{R}_-'
  \end{pmatrix},
\end{equation}
with $\mathcal{R}_{\pm}' \in \mathrm{SU}(R)$, such that
$\mathcal{R}' \in \mathrm{SU}(R) \times \mathrm{SU}(R)$. Note that matrices of
this form also obey condition (iii). To restrict this form further to
transformations $\mathcal{R}$ that obey also condition (ii), we factor out the
group of transformations $\mathcal{R}'$ that leave the replica-symmetric NLSM
manifold~\eqref{eq:Q-0} invariant:
\begin{equation}
  \left[ \mathcal{R}', M Q_0 M \right] = 0 \quad \Longleftrightarrow
  \quad \mathcal{R}_+' = \mathcal{R}_-'.
\end{equation}
Block-diagonal matrices $\mathcal{R}'$ that obey this condition form the group
$\mathcal{R}' \in \mathrm{SU}(R)$. The replicon manifold is thus
$\mathrm{SU}(R) \times \mathrm{SU}(R)/\mathrm{SU}(R) = \mathrm{SU}(R)$. An
explicit parametrization can be obtained by writing
\begin{equation}
  M Q_0 M =
  \begin{pmatrix}
    Q_{++} & Q_{+-} \\ Q_{-+} & Q_{--}
  \end{pmatrix},
\end{equation}
so that
\begin{equation}
  \label{eq:Q-U}
  Q = M \mathcal{R}' M Q_0 M \mathcal{R}^{\prime -1} M = M
  \begin{pmatrix}
    Q_{++} & Q_{+-} U \\ Q_{-+} U^{-1} & Q_{--}
  \end{pmatrix}
  M,
\end{equation}
with $U = \mathcal{R}_+' \mathcal{R}_-^{\prime - 1} \in \mathrm{SU}(R)$.

So far, we have focused on class AIII, to which our model belongs for
$\varphi \neq 0$. In the derivation of the NLSM in class BDI, realized for
$\varphi = 0$, the Keldysh action has to be symmetrized explicitly with respect
to PHS by introducing doubled spinors of fields, collecting the original
fermionic fields and their charge-conjugated partners~\cite{Poboiko2025}. After
this step, the construction of the NLSM target manifold is analogous to the one
presented above, yielding the replica-symmetric manifold
$\mathrm{O}(4)/\mathrm{U}(2) \simeq \mathrm{S}^2$ and the replicon manifold
$\mathrm{SU}(2 R)/\mathrm{Sp}(R)$, where $\mathrm{Sp}(R)$ is the compact
symplectic group~\cite{Poboiko2025, Starchl2025}.

We note in passing that the symmetries which we discuss here within the field
theory framework have clear analogs in the operator formalism and can also be
given an interpretation in terms of particle-number
conservation~\cite{Starchl2025}. In particular, the symmetry group
$\mathrm{U}(2)$ underlying the replica-symmetric manifold contains a subgroup
$\mathrm{U}(1) \times \mathrm{U}(1)$ that corresponds to a strong
$\mathrm{U}(1)$ symmetry in the operator formalism~\cite{Buca2012, Albert2014}.
This strong symmetry is associated with the conservation of the number of
particles in the system alone. In contrast, the symmetry group
$\mathrm{SU}(R) \times \mathrm{SU}(R)$ that underlies the replicon manifold and
corresponds to a strong $\mathrm{SU}(R)$ symmetry in the operator formalism is
present even under a weaker condition: The number of particles does not have to
be conserved within the system alone; rather, it has to be conserved within the
system and reservoirs that might be needed to realize more general measurement
processes~\cite{Starchl2025}.

\subsection{Nonlinear sigma model}
\label{sec:NLSM-action-RG}

The expansion of the action in fluctuations around a given saddle point, which
forms the basis of the Gaussian theory discussed in
Sec.~\ref{sec:gaussian-theory}, is controlled for $\gamma \ll J$, provided that
the fluctuations remain sufficiently small~\cite{Poboiko2023}. This condition is
satisfied for massive fluctuations.  In contrast, massless modes exhibit strong
fluctuations that invalidate the Gaussian approximation at large scales. These
strong fluctuations occur within the NLSM manifold and are parameterized by
Eqs.~\eqref{eq:Q-0} and~\eqref{eq:Q-U}. A standard procedure to derive the NLSM
action for the massless modes is to substitute this parameterization into the
action~\eqref{eq:S-G-Sigma}, take the spatial continuum limit in which the
lattice index $l$ is replaced by a continuous variable $x$, and perform a
gradient expansion~\cite{Poboiko2025}. The resulting NLSM extends the Gaussian
theory of Sec.~\ref{sec:gaussian-theory} by avoiding an expansion in the
massless modes; however, fluctuations of the massive modes are then
neglected. For our purposes, this is insufficient, as it produces an NLSM that
fails to capture the dependence of the physics on the unraveling phase
$\varphi$. To understand this, note that the action~\eqref{eq:S-G-Sigma}
consists of two contributions given in Eqs.~\eqref{eq:S-0-G-Sigma}
and~\eqref{eq:L-M-G}. Only the former, which is independent of $\varphi$,
contributes nontrivially to the NLSM action. The latter depends on $\varphi$,
but its contribution vanishes on the NLSM manifold and in the limit $R \to
1$. Consequently, the resulting NLSM is independent of $\varphi$.

In principle, the $\varphi$-dependence of the NLSM coupling constants can be
restored by generalizing Eq.~\eqref{eq:G-Sigma-Q} to include massive
fluctuations outside the NLSM manifold, as incorporated in the Gaussian
ansatz~\eqref{eq:G-Sigma-fluctuations}. Integrating out these massive
fluctuations within the Gaussian approximation---which remains controlled for
massive modes on all scales---then leads to a renormalization of the NLSM
coupling constants. A more direct alternative is to start from the general form
of the NLSM obtained via the gradient expansion of Eq.~\eqref{eq:S-0-G-Sigma},
leaving the coefficients undetermined, then linearize the NLSM action and match
it to the Gaussian theory after the massive fluctuations have been integrated
out in the latter. Following this approach, we find that the NLSM action splits
into two contributions,
\begin{equation}
  S[Q] = S_0[Q_0] + S_R[U, Q_0],
\end{equation}
that describe, respectively, the replica-symmetric and replicon sectors of the
theory. As noted in Sec.~\ref{sec:replica-keldysh-action}, for $R = 1$, the
measurement Lagrangian~\eqref{eq:L-M-G-R=1} coincides with the one for random
projective measurements of occupation numbers~\cite{Poboiko2023, Poboiko2025},
up to a factor of $1/2$ in the definition of $\gamma$. Therefore, in the
replica-symmetric sector of the theory, the Lagrangian is
\begin{equation}
  \imag \mathcal{L}_0[Q_0] = \frac{1}{2} \tr \! \left[ \sigma_z
    \mathcal{R}_0^{-1} \partial_t \mathcal{R}_0 - \frac{\nu}{4}
    \left( \partial_x Q_0 \right)^2 \right],
\end{equation}
where $\nu = J^2/\gamma$. For class AIII, we consider here transformations
$\mathcal{R}_0 \in \mathrm{U}(2)/\mathrm{U}(1) \times \mathrm{U}(1)$ and $Q_0$
is given in Eq.~\eqref{eq:Q-0}. The NLSM on the replica-symmetric manifold
describes diffusive dynamics on scales larger than $l_0$
Eq.~\eqref{eq:u-l-0}~\cite{Poboiko2023, Poboiko2025}.

The NLSM Lagrangian in the replicon sector is
\begin{equation}
  \label{eq:L-replicon}
  \imag \mathcal{L}_R[U, Q_0] = - \frac{g_{\varphi}}{2} \trR \! \left(
    \frac{1}{v_{\varphi}} \partial_t U^{-1} \partial_t U +
    v_{\varphi} \partial_x U^{-1} \partial_x U \right),
\end{equation}
with, for class AIII, $U \in \mathrm{SU}(R)$ as given below
Eq.~\eqref{eq:Q-U}. The parameters appearing in the NLSM Lagrangian are
\begin{equation}
  g_{\varphi} = \frac{l_0 \rho_0 \left( 1 - \rho_0 \right)}{\cos(\varphi)},
  \quad v_{\varphi} = 4 J \cos(\varphi) \sqrt{\rho_0 \left( 1 - \rho_0 \right)},
\end{equation}
where the replica-symmetric density is defined as in
Eq.~\eqref{eq:rho-classical}:
\begin{equation}
  \label{eq:rho-0}
  \rho_0 = \frac{1}{4} \trK \! \left( 1 - \sigma_x Q_0 \right).
\end{equation}
In the limit $\gamma \ll J$, in which the derivation of the NLSM is valid, we
can neglect fluctuations in the replica-symmetric sector and replace the density
by its value $\rho_0 = n$ at the saddle point $Q_0 = \Lambda$. Then, the bare
coupling constant of the NLSM reduces to the one of the Gaussian
theory~\eqref{eq:g-phi-0}, $g_{\varphi} = g_{\varphi, 0}$. Note that the
coupling constant diverges for $\varphi \to \pi/2$. We comment on the NLSM
description for the case $\varphi = \pi/2$ in Sec.~\ref{sec:unitary-unraveling}
below.

The NLSM for symmetry class BDI takes the same form as the one for class AIII,
but with matrix fields $\mathcal{R}_0$ and $U$ belonging to different manifolds
according to our discussion in Sec.~\ref{sec:target-manifold}. As a
result, the one-loop RG flow equations for the two symmetry classes differ only
by a numerical prefactor~\cite{Poboiko2023, Poboiko2025}. The RG flow of the
NLSM coupling constant $g_{\varphi}$ reads
\begin{equation}
  \label{eq:g-phi-RG}
  g_{\varphi} = g_{\varphi, 0} - \frac{1}{4 \pi \beta} \ln(l/l_0),
\end{equation}
where $\beta = 1$ in class AIII for $\varphi \neq 0$, and $\beta = 1/2$ in class
BDI for $\varphi = 0$. For $\gamma \ll J$, the bare coupling~\eqref{eq:g-phi-0}
$g_{\varphi, 0} \sim l_0 \sim J/\gamma$ is large, and Eq.~\eqref{eq:g-phi-RG}
describes the slow RG flow toward the strong-coupling regime of the NLSM where
$g_{\varphi} \lesssim 1$. The strong-coupling regime is reached beyond the scale
\begin{equation}
  \label{eq:l-phi-star}
  l_{\varphi, *} = l_0 \e^{4 \pi \beta g_{\varphi, 0}} = l_0 \exp \! \left[ \frac{8
      \pi \beta J n \left( 1 - n \right)}{\gamma \cos(\varphi)} \right].
\end{equation}

As stated at the end of Sec.~\ref{sec:entanglement-entropy-Gaussian}, the
renormalization of $g_{\varphi}$ determines the behavior of the effective
central charge on large scales. Therefore, the flow of $g_{\varphi}$ to small
values beyond the scale $l_{\varphi, *}$ implies that the logarithmic growth of
the entanglement entropy $S_{\ell}$ described by Eq.~\eqref{eq:S-l-asymptotic}
saturates, and that $S_{\ell}$ shows area-law behavior for
$\ell \gtrsim l_{\varphi, *}$. This in turn indicates that there is no
measurement- or unraveling-induced phase transition between a critical and an
area-law phase in our model. Instead, the entanglement entropy obeys an area law
beyond the scale $l_{\varphi, *}$ for all values of $\gamma$ and
$0 \leq \varphi < \pi/2$. However, $l_{\varphi, *}$ is exponentially large in
$J/\left[ \gamma \cos(\varphi) \right]$, and therefore we do not observe
area-law entanglement in our numerical simulations, which we discuss in detail
in Sec.~\ref{sec:numerics-vs-theroy} below, for large values of
$J/\left[ \gamma \cos(\varphi) \right]$, meaning that either $\gamma$ is small
or $\varphi$ is close to $\pi/2$. Finally, note that the exponential dependence
implies that $l_{\varphi, *}$ drops sharply upon decreasing
$J/\left[ \gamma \cos(\varphi) \right]$ by either increasing $\gamma$ or
decreasing $\varphi$. In numerics, this sharp drop might be misinterpreted as a
measurement- or unraveling-induced phase transition.

\subsection{RG-corrected Gaussian theory}
\label{sec:rg-corrected-gauss}

Because area-law behavior is expected only at exponentially large scales, making
it difficult to capture numerically, it is useful to identify signatures of the
RG flow~\eqref{eq:g-phi-RG} that are visible already at shorter scales. To this
end, we incorporate the RG correction Eq.~\eqref{eq:g-phi-RG} into the Gaussian
correlation function~\eqref{eq:C-q-Gaussian-bulk}~\cite{Poboiko2023,
  Starchl2025}. This is done by replacing the bare NLSM coupling
$g_{\varphi, 0}$ by its renormalized value~\eqref{eq:g-phi-RG}, with the
momentum at which the correlation function is evaluated providing the IR cutoff
of the RG flow according to $l = 1/q$.  Furthermore, in
Eq.~\eqref{eq:C-q-Gaussian-bulk} we replace $\tilde{c}(u)/u \sim 1$ by its
asymptotic value for small momenta given in
Eq.~\eqref{eq:c-tilde-asymptotic}. We thus obtain
\begin{equation}
  \label{eq:C-q-RG-corrected}
  C_q \sim C_{q, 0} + \frac{q}{4 \pi \beta} \ln(q l_0),
\end{equation}
where $C_{q, 0} \sim g_{\varphi, 0} q$ describes the asymptotic behavior of the
Gaussian correlation function~\eqref{eq:C-q-Gaussian-bulk} for $q \to 0$. This
form suggests that the ``weak-localization correction,'' defined as the
difference between the exact connected density correlation function and the
Gaussian approximation, provides a direct and numerically accessible probe of
the RG flow~\eqref{eq:g-phi-RG}~\cite{Poboiko2023, Poboiko2025}:
\begin{equation}
  \label{eq:weak-localization-correction}
  \frac{\delta C_q}{q} = \frac{C_q - C_{q, 0}}{q} = \frac{1}{4 \pi \beta} \ln(q
  l_0).
\end{equation}
The slope of $\delta C_q/q$ as a function of $\ln(q l_0)$ is universal, and its
value distinguishes between symmetry classes AIII with $\beta = 1$ for
$\varphi \neq 0$ and BDI with $\beta = 1/2$ for $\varphi = 0$.  Note, however,
that the weak-localization correction acquires a nonuniversal shift that depends
on the UV cutoff, which we have identified with $1/l_0$ in
Eq.~\eqref{eq:g-phi-RG}.

For any value of $\varphi < \pi/2$, the functional form of the renormalized
correlation function~\eqref{eq:C-q-RG-corrected} is the same as that obtained
for different models in Refs.~\cite{Poboiko2023, Starchl2025}. Therefore, one
might expect the same qualitative behavior to apply: Signatures of emergent
conformal invariance as characteristic for a critical phase~\cite{Alberton2021,
  Eissler2025} can be observed on a finite critical range of length scales,
bounded from below by $l_0 \sim \gamma^{-1}$ and from above by a scale
$l_c \sim \gamma^{-2}$. These signatures include algebraic decay of the density
correlation function, $C_l \sim l^{-2}$, and logarithmic growth of the
entanglement entropy, $S_{\ell} \sim \ln(\ell)$. Note that these properties are
also predicted by the Gaussian theory in
Eqs.~\eqref{eq:C-l-Gaussian-bulk-approximation} and~\eqref{eq:S-l-asymptotic},
respectively. However, the Gaussian value of the central charge,
Eq.~\eqref{eq:effective-central-charge-Gaussian}, applies only asymptotically
for $\gamma \to 0$ and $\ell \to \infty$. For any finite value of $\gamma$,
approximately logarithmic growth of the entanglement entropy occurs not for
$\ell \to \infty$ but on intermediate scales where $c_{\ell}$ is approximately
constant, in the vicinity of the maximum of $c_{\ell}$ at
$\ell = l_m \sim \gamma^{-3/2}$.  The scaling of $l_m$ with $\gamma$ ensures
that this scale is within the critical range, $l_0 < l_m < l_c$. Further
signatures of conformal invariance, such as a universal functional form of the
mutual information, are reproduced with the value of the effective central
charge $c_{\ell}$ at its maximum~\cite{Starchl2025}.

\section{Numerical validation of field-theory predictions}
\label{sec:numerics-vs-theroy}

To validate our analytical findings, we compare them to direct numerical
simulations of Eq.~\eqref{eq:sse}. We anticipate that the expectations
formulated at the end of Sec.~\ref{sec:rg-corrected-gauss} are borne out by our
numerics for small values of $\varphi$; for large values of $\varphi$, we still
find the weak-localization correction as in
Eq.~\eqref{eq:weak-localization-correction}, indicating that the model shows
area-law entanglement on large scales. However, we observe additional structure
in short-scale nonuniversal regime $l \lesssim l_0$ that is described by the
Gaussian theory. This additional structure masks the scaling behavior
$l_c \sim \gamma^{-2}$ and $l_m \sim \gamma^{-3/2}$. Before we discuss these
results in detail, we briefly summarize the numerical methods we use.

\subsection{Numerical methods}
\label{sec:numerical-methods}

\begin{figure*}
    \centering    
    \includegraphics{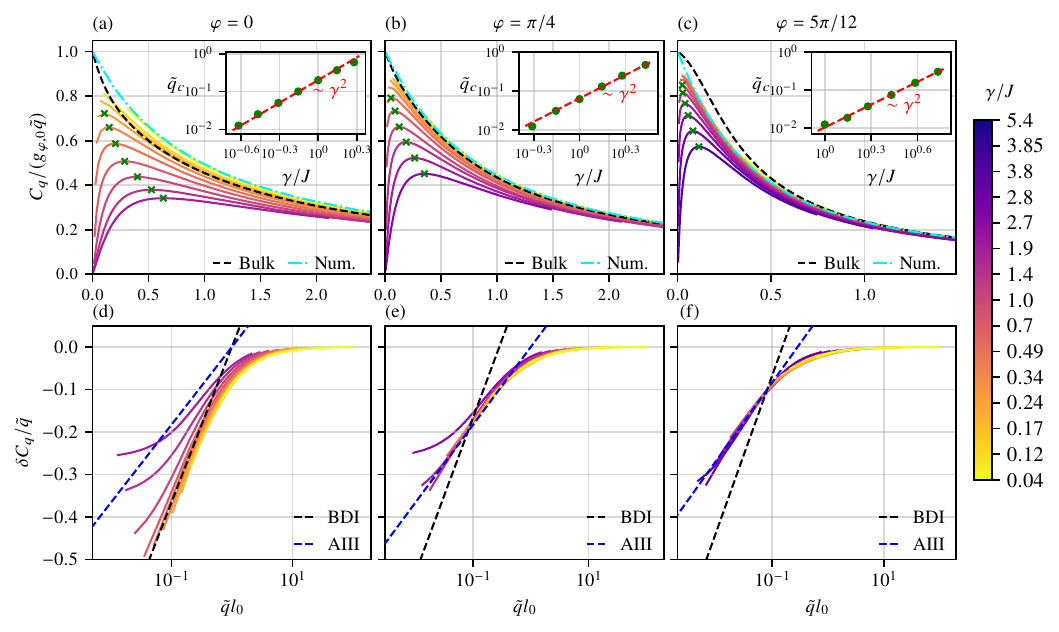}
    \caption{(a--c) Rescaled density correlation
      function~\eqref{eq:connected-density-correlation-function} in momentum
      space and (d--f) weak-localization
      correction~\eqref{eq:weak-localization-correction} for (a,
      d)~$\varphi = 0$, (b, e)~$\varphi = \pi/4$, and (c,
      f)~$\varphi = 5 \pi/12$. (a--c) The Gaussian correlation function (cyan
      dash-dotted line: full numerical
      solution~\eqref{eq:C-l-Gaussian-full-numerical}; black dashed line: bulk
      approximation~\eqref{eq:C-q-Gaussian-bulk}) decreases monotonically
      starting from $C_q/(g_{\varphi, 0} q) \to 1$ for $\tilde{q} l_0 \to 0$. In
      contrast, the numerical data attain a maximum at $q_c$ and decrease for
      $\tilde{q} l_0 \to 0$. This decrease becomes more pronounced for
      increasing values of $\gamma/J$. Inset in (a--c): The position of the
      maximum scales as $q_c \sim \gamma^2$. (d--f) The weak-localization
      correction agrees well with the one-loop RG
      result~\eqref{eq:weak-localization-correction}. In particular, the
      analytically predicted difference in the slope by a factor of two between
      symmetry classes BDI for $\varphi = 0$ and AIII for $\varphi \neq 0$ is
      borne out by the data. Note that the analytical
      result~\eqref{eq:weak-localization-correction} is shifted for better
      visual comparison with the numerical data. Such a nonuniversal shift
      corresponds to a different value of the UV cutoff in the RG
      flow~\eqref{eq:g-phi-RG} and does not affect the universal slope of the
      weak-localization correction.}
    \label{fig:Cqg0_vs_qT_plot}
\end{figure*}

To numerically simulate the dynamics generated by the stochastic Schr\"odinger
equation~\eqref{eq:sse}, we adapt the approach first introduced by Cao et
al.~\cite{Cao2019}. In all of our simulations, the initial state is chosen as
\begin{equation}
  \label{eq:ket-psi-CDW}
  \ket{\psi_0} = \prod_{l = 1}^{L/2} \hat{\psi}_{2 l - 1}^{\dagger} \ket{0},
\end{equation}
where $\ket{0}$ is the vacuum state. The number of particles is thus $N = L/2$.
Both the number of particles and the Gaussianity of the initial state are
preserved by Eq.~\eqref{eq:sse}. Therefore, at all times $t$ the state can be
written as
\begin{equation}
  \label{eq:psi-U}
  \ket{\psi(t)} = \prod_{n = 1}^N \left[\sum_{l = 1}^L
    U_{l, n}(t) \hat{\psi}_l^{\dagger}\right] \ket{0},
\end{equation}
where the $L \times N$ matrix $U(t)$ is an isometry, $U^{\dagger}(t) U(t) = 1$,
which ensures the normalization of the state. Since the quantum state
$\ket{\psi(t)}$ is fully determined by $U(t)$, the stochastic Schr\"odinger
equation~\eqref{eq:sse} can be recast as an evolution equation for this
matrix. To obtain this evolution equation, we insert Eq.~\eqref{eq:psi-U} in
Eq.~\eqref{eq:sse}, commute the operators $\hat{H}$ and $\hat{n}_l$ to the right
using $\hat{H} \ket{0} = \hat{n}_l \ket{0} = 0$, and exponentiate the generator
of the evolution of $U(t)$ using the It\^o rule in Eq.~\eqref{eq:Ito-rules},
$\diff W_l(t) \diff W_{l'}(t) = \delta_{l, l'} \diff t$. We thus obtain
\begin{equation}  
  U(t + \diff t) = M \e^{-\imag H \diff t} U(t),
\end{equation}
where $H$ is the Hamiltonian matrix~\eqref{eq:Hamiltonian-matrix} and the
diagonal matrix describing the measurement process is given by
\begin{multline}  
  M_{l, l'} = \delta_{l, l'} \exp \! \left\{ \frac{\gamma}{2} \left[ 2 \langle
      \hat{n}_l(t) \rangle - 1 \right] \left( 1 + \e^{2 \imag \varphi} \right)
    \diff t \right. \\ \left. \vphantom{\frac{\gamma}{2}} + \sqrt{\gamma}
    \e^{\imag \varphi} \diff W_l(t) \right\}.
\end{multline}
To calculate expectation values such as $\langle \hat{n}_l(t) \rangle$ in the
time-evolved state $\ket{\psi(t)}$, we use Wick's theorem and the fact that the
$L \times L$ single-particle density matrix is related to $U(t)$ via
\begin{equation}
  \label{eq:single-particle-density-matrix}
  D_{l, l'}^{}(t) = \bra{\psi(t)} \hat{\psi}_{l'}^{\dagger} \hat{\psi}_l^{}
  \ket{\psi(t)} = \sum_{n = 1}^N U_{l, n}^{}(t) U_{l', n}^{*}(t).
\end{equation}
We simulated the time evolution of $U(t)$ for a system of $L = 1000$ lattice
sites and using a time step $J \Delta t = 0.05$, which has been shown to yield
sufficient accuracy in the limiting cases $\varphi = 0$ and
$\varphi = \pi/2$~\cite{Cao2019}. After each time step, a QR decomposition was
applied to restore the isometric property of $U(t+\diff t)$~\cite{Cao2019}. For
each choice of the parameters $\gamma$ and $\varphi$, we simulated
$N_{\mathrm{traj}} = 160$ trajectories.

\subsection{Connected density correlation function}
\label{sec:conn-dens-corr}

\begin{figure*}
  \centering
  \includegraphics{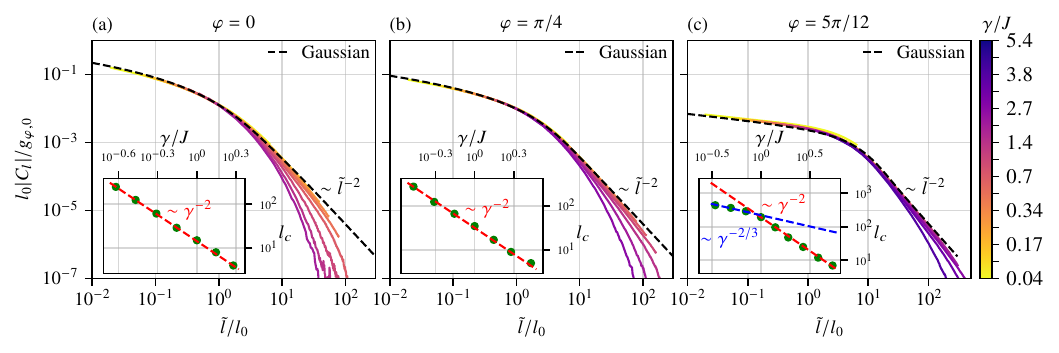}
  \caption{Rescaled density correlation
    function~\eqref{eq:connected-density-correlation-function} for
    (a)~$\varphi = 0$, (b)~$\varphi = \pi/4$, and (c)~$\varphi = 5 \pi/12$. On
    short scales $\tilde{l} \lesssim l_0$, the numerical data agree well with
    the Gaussian result (black dashed line: bulk
    approximation~\eqref{eq:C-q-Gaussian-bulk}). Significant deviations occur
    for $\tilde{l} \gg l_0$, where the Gaussian result decays as a power law,
    $\abs{C_l} \sim \tilde{l}^{-2}$, while the numerical data exhibit faster
    decay. Inset: The crossover scale at which deviations from the Gaussian
    result become significant scales as $l_c \sim \gamma^{-2}$. For
    $\varphi = 5 \pi/12$, this scaling changes to $l_c \sim \gamma^{-2/3}$ at
    small values of $\gamma/J$.}
    \label{fig:Clg0_vs_l_plot}
\end{figure*}

We first consider the connected density correlation
function~\eqref{eq:C-l-l-prime-t}. In the steady state, the correlation function
is time-independent and translationally invariant, $C_{l, l'}(t) = C_{l - l'}$.
Using Wick's theorem to express the correlation function in terms of the
single-particle density matrix~\eqref{eq:single-particle-density-matrix}, we
obtain
\begin{equation}
  \label{eq:connected-density-correlation-function}
  C_{l - l'} = \overline{\left\langle \hat{n}_l \hat{n}_{l'} \right\rangle} -
  \overline{\left\langle \hat{n}_l \rangle \langle \hat{n}_{l'}
    \right\rangle} = \delta_{l, l'}/2
  - \overline{\lvert D_{l, l'} \rvert^2}.
\end{equation}
For our numerical results, the overline denotes averaging over both quantum
trajectories and lattice sites $l$ and $l'$, keeping the distance $l - l'$
constant.

Figure~\ref{fig:Cqg0_vs_qT_plot}(a--c) shows the rescaled density correlation
function in momentum space, $C_q/(g_{\varphi, 0} \tilde{q})$, as a function of
$\tilde{q} l_0$ for various values of $\varphi$ and $\gamma/J$. Here,
$\tilde{q} = 2 \sin(q/2)$ takes the finite lattice spacing into
account~\cite{Poboiko2023}. For comparison, also the correlation functions
obtained from the Gaussian theory, both the bulk
approximation~\eqref{eq:C-q-Gaussian-bulk} and the full numerical
solution~\eqref{eq:C-l-Gaussian-full-numerical}, are shown in the figure.

As expected, the numerical data are close to the Gaussian results for small
values of $\gamma/J$ and for intermediate to large values of $q l_0$. However,
we observe qualitatively different behavior at low momenta: The asymptotic
behavior~\eqref{eq:c-tilde-asymptotic} of the Gaussian correlation
function---recall that the bulk approximation and the full numerical solution
have the same asymptotic behavior---implies that $C_q/(g_{\varphi, 0} q) \to 1$
for $q l_0 \to 0$. In contrast, strong nonlinear fluctuations renormalize
$C_q/(g_{\varphi, 0} q)$ to small values for $q l_0 \to 0$, as described by the
RG correction in Eq.~\eqref{eq:C-q-RG-corrected}. The numerical data clearly
follow this trend. We identify the crossover scale $q_c$ between Gaussian
behavior at large momenta and the fluctuation-dominated regime at small momenta
with the position of the maximum of $C_q/(g_{\varphi, 0} q)$, which can be
obtained analytically from Eq.~\eqref{eq:C-q-RG-corrected} by including the
quadratic order in the expansion
Eq.~\eqref{eq:c-tilde-asymptotic}~\cite{Starchl2025}. As explained below
Eq.~\eqref{eq:c-tilde-asymptotic}, the coefficient of the quadratic order is not
correctly reproduced by the bulk approximation. This coefficient affects the
dependence of $q_c$ on $\varphi$. However, precise knowledge of the coefficient
is not necessary to determine the scaling $q_c \sim \gamma^2$. The numerical
data, shown in the inset in Fig.~\ref{fig:Cqg0_vs_qT_plot}(a--c), are in good
agreement with this scaling.

Figure~\ref{fig:Cqg0_vs_qT_plot}(d--f) shows the weak-localization
correction~\eqref{eq:weak-localization-correction}, defined as the difference
between the numerical data for $C_q$ and the Gaussian correlation
function~\eqref{eq:C-l-Gaussian-full-numerical}. We observe good agreement with
the one-loop RG result~\eqref{eq:weak-localization-correction}. In particular,
the distinction between symmetry classes BDI for $\varphi = 0$ and AIII for
$\varphi \neq 0$ is well supported by the data. Values of $\varphi$ other than
those shown in Fig.~\ref{fig:Cqg0_vs_qT_plot} yield similar results, with a
crossover from the behavior expected for symmetry classes BDI and AIII at small
values of $\varphi$~\cite{Niederegger2025}. To obtain these results, we used the
full numerical solution for the Gaussian correlation
function~\eqref{eq:C-l-Gaussian-full-numerical}. The bulk approximation is not
sufficiently accurate for this purpose.

The agreement between the numerical results for the weak-localization correction
and the analytical prediction based on the NLSM provides strong evidence for the
absence of an entanglement transition in our model. However, as was found in a
variety of similar models, signatures that are characteristic for a critical
phase can be observed on intermediate length scales~\cite{Alberton2021,
  Eissler2025, Poboiko2023, Starchl2025}. As examples for such signatures, we
focus here on algebraic decay of the density correlation function and
logarithmic growth of the entanglement entropy.

Figure~\ref{fig:Clg0_vs_l_plot} shows the rescaled connected density correlation
function~\eqref{eq:connected-density-correlation-function} as a function of the
chord length $\tilde{\ell} = \left( L/\pi \right) \sin(\pi\ell/L)$, which takes
both the finite size of the system and the periodic boundary conditions into
account~\cite{DiFrancesco1997}. For comparison, the black dashed line shows the
bulk approximation for the Gaussian correlation function, obtained by
transforming Eq.~\eqref{eq:C-q-Gaussian-bulk} to real space. On intermediate
scales, $l_0 < l < l_c$, the numerical data exhibit approximately algebraic
decay, $C_l \sim \tilde{l}^{-2}$, as characteristic for a 1D CFT. Following
Ref.~\cite{Starchl2025}, we identify $l_c$ with the scale beyond which the
deviation between the numerical data and a tangent to the data
$\sim \tilde{l}^{-2}$ exceeds $10 \, \%$. As discussed in more detail below,
$l_c$ depends algebraically on $\gamma$.  A key practical consequence of this
algebraic scaling is that $l_c$ remains smaller than the system size $L = 1000$
in our numerics and is therefore accessible over a broad range of $\gamma/J$. By
contrast, the scale $l_{\varphi,*}$, which signals the onset of the area-law
regime, increases exponentially with $J/[\gamma \cos(\varphi)]$ according to
Eq.~\eqref{eq:l-phi-star}. As already noted in Sec.~\ref{sec:previous-work},
resolving this scale in numerical simulations is therefore extremely
challenging~\cite{Fan2025}. More specifically, for the parameters used in
Fig.~\ref{fig:Clg0_vs_l_plot}(a)--(c), corresponding to
$\varphi = 0, \pi/4, 5\pi/12$, the scale $l_{\varphi,*}$ exceeds the system size
$L$ for $\gamma/J \lesssim 0.6, 1.4, 3.3$, respectively. For the smallest value
considered in our numerics, $\gamma/J = 0.04$, we estimate $l_{\varphi,*}$ to be
at least tens of orders of magnitude larger than $L$, placing it well beyond the
reach of our numerical approach.

The insets of Fig.~\ref{fig:Clg0_vs_l_plot} show good agreement of the numerical
data for $l_c$ with the expected scaling $l_c \sim q_c^{-1} \sim \gamma^{-2}$
for $\varphi = 0$ and $\varphi = \pi/4$.  However, for $\varphi = 5 \pi/12$, we
observe a different scaling $l_c \sim \gamma^{-2/3}$ at small values of
$\gamma/J$. We attribute the modified scaling of $l_c$ to a change in the shape
of the correlation function in the nonuniversal regime $l \lesssim l_0$. In this
regime, the numerical data agree well with the Gaussian approximation, and the
change in the shape can be seen best in the latter: For small values of
$\varphi$, the Gaussian correlation function is concave; however, close
inspection of Fig.~\ref{fig:Clg0_vs_l_plot}(c) reveals that for
$\varphi = 5 \pi/12$, the Gaussian correlation function exhibits a ``dent''
between $\tilde{l}/l_0 \approx 10$ and $\tilde{l}/l_0 \approx 10^2$. Recall that
our method to determine $l_c$ involves drawing a tangent to the correlation
function. Clearly, the point at which the tangent meets the correlation function
is affected by this change of shape. This suggests that the change of shape is
also the cause for the modified scaling behavior of $l_c$. We did not attempt to
determine precisely the value of $\varphi$ at which the Gaussian correlation
function ceases to be concave; however, the dent is there also for
$\varphi = \pi/3$, which indicates that it appears first between
$\varphi = \pi/4$ and $\varphi = \pi/3$~\cite{Niederegger2025}.

Note that in Fig.~\ref{fig:Cqg0_vs_qT_plot}(c), the deviation of the data point
for the smallest value of $\gamma/J$ from $q_c \sim \gamma^2$ is compatible with
the onset of the scaling $q_c \sim l_c^{-1} \sim \gamma^{2/3}$. However, we have
defined $q_c$ as the position of the maximum of $C_q/(g_{\varphi, 0} q)$; for
the smallest values of $\gamma/J$, we do not find a maximum within the range of
values of $\tilde{q} l_0$, which is bounded from below by the inverse system
size. Therefore, conclusively demonstrating the scaling $q_c \sim \gamma^{2/3}$
would require larger system sizes.

As we discuss next, we find a similar modification of scaling with $\gamma$ in
the effective scale-dependent central
charge~\eqref{eq:effective-central-charge}. In this case, the origin of the
modification is particularly clear, and the modified scaling can be read off
directly from the Gaussian result.

\subsection{Entanglement entropy and effective central charge}

\begin{figure*}
  \centering
  \includegraphics{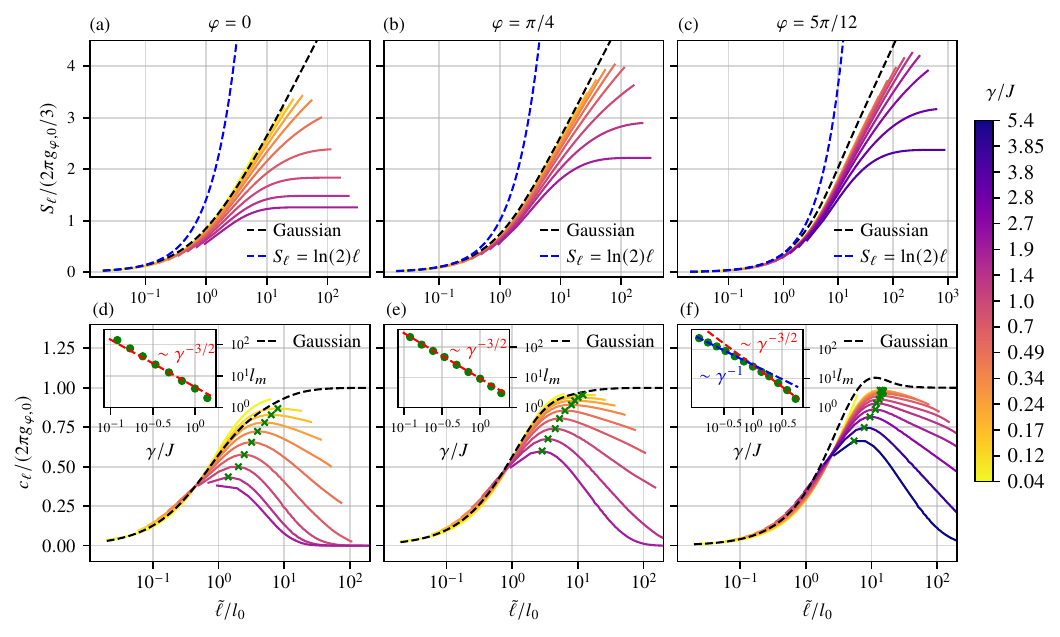}
  \caption{(a--c) Rescaled entanglement entropy and (d--f) rescaled
    scale-dependent effective central charge for (a, d)~$\varphi = 0$, (b,
    e)~$\varphi = \pi/4$, and (c, g)~$\varphi = 5 \pi/12$. (a--c) For small
    values of $\gamma/J$, volume-law scaling of the entanglement entropy (blue
    dashed line) on short scales $\tilde{\ell} \lesssim l_0$ crosses over to
    apparently logarithmic growth on large scales. At larger values of
    $\gamma/J$, the data exhibit area-law scaling. (d--f) Unlike the Gaussian
    result (black dashed line), the numerical data for the effective central
    charge do not saturate to a finite value as $\tilde{\ell}/l_0 \to \infty$.
    This implies that the growth of the entanglement entropy is, in fact, not
    logarithmic on large scales. Inset: The effective central charge attains its
    maximum at $l_m \sim \gamma^{-3/2}$. For $\varphi = 5 \pi/12$, the scaling
    is modified to $l_m \sim \gamma^{-1}$ at small values of $\gamma/J$.}
  \label{fig:entanglement-entropy-central-charge}
\end{figure*}

To calculate the entanglement entropy~\eqref{eq:entanglement-entropy}
numerically, we use the following relation for Gaussian states:
\begin{equation}
  \svn{\ell} = - \sum_{l = 1}^{\ell} \overline{\left[ \lambda_l \ln(\lambda_l) +
      \left( 1 - \lambda_l \right) \ln(1 - \lambda_l) \right]},
\end{equation}
where $\lambda_l$ are the eigenvalues of the single-particle density
matrix~\eqref{eq:single-particle-density-matrix}, restricted to a subsystem of
size $\ell$; that is,
$D_{\ell} = (D_{l, l'})_{l, l' = 1}^{\ell}$~\cite{Peschel2009}. As in
Eq.~\eqref{eq:connected-density-correlation-function}, the overline denotes
spatial and trajectory averaging.

Figure~\ref{fig:entanglement-entropy-central-charge}(a--c) shows the
entanglement entropy as a function of subsystem size
$\ell \in \{ 1, \dotsc, L/2 \}$ for three different values of $\varphi$. In each
case, the numerical data exhibit the same qualitative behavior: For small
$\gamma/J$, the data closely follow the Gaussian curve, obtained by numerically
integrating Eq.~\eqref{eq:entanglement-entropy-correlation-function} with the
Gaussian correlation function in bulk
approximation~\eqref{eq:C-q-Gaussian-bulk}. In particular, the data exhibit
volume-law scaling on short scales $\ell \lesssim l_0$, and apparently
logarithmic growth on large scales. For increasing $\gamma/J$, the large-scale
behavior obeys an area law. The NLSM description implies that area-law scaling
occurs for all values of $\gamma$ and $\varphi$, but only beyond the
exponentially large scale $l_{\varphi, *}$~\eqref{eq:l-phi-star}. This is not
observable for small values of $\gamma/J$ and the system size of $L = 1000$
which we have chosen in our numerical simulations. Nevertheless, as we discuss
next, the onset of the crossover toward area-law scaling is clearly visible in
the scale-dependent effective central
charge~\eqref{eq:effective-central-charge}.

The effective central charge is shown in
Fig.~\ref{fig:entanglement-entropy-central-charge}(d--f). Again, for small
$\gamma/J$, the numerical data are close to the Gaussian result. However, for
the central charge, quantitative corrections to the bulk approximation appear to
be more pronounced. In particular, for small values of $\varphi$, there are
clear quantitative discrepancies between the Gaussian result and the numerical
data. Aside from this quantitative difference, there is a crucial
\emph{qualitative} difference between the Gaussian theory and the full-scale
numerics: On large scales, the Gaussian result approaches the constant value
given in Eq.~\eqref{eq:effective-central-charge-Gaussian}; a constant value of
the scale-dependent effective central charge implies logarithmic growth of the
entanglement entropy~\eqref{eq:S-l-asymptotic}. In contrast, for all but the
smallest value of $\gamma/J$ included in
Fig.~\ref{fig:entanglement-entropy-central-charge}, the central charge obtained
from our numerical simulations is approximately constant only in the vicinity of
its maximum, and it falls off on large scales. This is a clear signature of the
crossover from logarithmic growth towards area-law scaling of the entanglement
entropy. However, within the limited range of subsystem sizes in our
simulations, the central charge reaches zero---or, equivalently, the
entanglement entropy exhibits area-law scaling---only for large $\gamma/J$.

As stated above, we observe a maximum of the effective central charge for all
but the smallest value of $\gamma/J$. The absence of a maximum for small values
of $\gamma/J$ is due to the limited system size in our numerical
simulations. This conclusion is supported by the data for the position $l_m$ of
the maximum of $c_{\ell}$ shown in the inset in
Fig.~\ref{fig:entanglement-entropy-central-charge}(d--f). For $\varphi = 0$ and
$\varphi = \pi/4$, the data are compatible with the expected scaling
$l_m \sim \gamma^{-3/2}$~\cite{Starchl2025}. Therefore, for a given system size
$L$, the maximum is not observable if $\gamma/J$ is too small. Note that the
scaling $l_m \sim \gamma^{-3/2}$ places the maximum of the central charge and
thus the region of approximately logarithmic growth of the entanglement entropy
within the critical range, which is bounded from above and below by
$l_0 \sim \gamma^{-1}$ and $l_c \sim \gamma^{-2}$, respectively.

For $\varphi = 0$ and $\varphi = \pi/4$, the Gaussian result for $c_{\ell}$
grows monotonically, and the maximum of $c_{\ell}$ results from the approach to
the asymptotic value given in Eq.~\eqref{eq:effective-central-charge-Gaussian}
being counteracted by the renormalization of $g_{\varphi, 0}$ according to
Eq.~\eqref{eq:g-phi-RG}. In other words, the scaling $l_m \sim \gamma^{-3/2}$ is
determined by universal long-wavelength properties of the theory. However, for
$\varphi = 5 \pi/12$, the data in
Fig.~\ref{fig:entanglement-entropy-central-charge}(f) show the modified scaling
$l_m \sim \gamma^{-1}$ at the smallest values of $\gamma/J$---similarly to the
modified scaling of $l_c$ in Fig.~\ref{fig:Clg0_vs_l_plot}(f); and the modified
scaling $l_m \sim \gamma^{-1}$ for $\varphi = 5 \pi/12$ is a \emph{nonuniversal}
feature. To see that this is the case, note that as expected for very small
$\gamma/J$ and on short and intermediate length scales, the numerical data for
$c_{\ell}$ agree reasonably well with the Gaussian theory. But for
$\varphi = 5 \pi/12$, the Gaussian theory $c_{\ell}$ exhibits a maximum at a
fixed value of $\ell/l_0$, such that $l_m \sim l_0 \sim \gamma^{-1}$. As stated
above, the Gaussian result for $c_{\ell}$ shown in
Fig.~\ref{fig:entanglement-entropy-central-charge} is derived from the bulk
approximation, and we expect the full numerical solution of the Gaussian theory
to exhibit quantitative corrections. However, the key new qualitative feature
appearing for increasing $\varphi$, which is the maximum of $c_{\ell}$, appears
to be reproduced correctly also in the bulk approximation. We observe this
maximum also for $\varphi = \pi/3$ but not for $\varphi = \pi/4$, indicating
that upon increasing $\varphi$ the maximum appears first between
$\varphi = \pi/4$ and $\varphi = \pi/3$, as is the case for the dent in the
connected density correlation function discussed in
Sec.~\ref{sec:conn-dens-corr}~\cite{Niederegger2025}.

\begin{figure}
    \centering
    \includegraphics{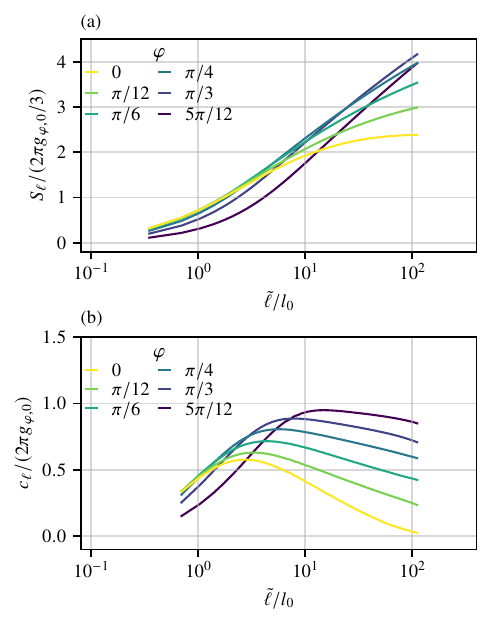}
    \caption{(a) Rescaled entanglement entropy and (b) rescaled effective
      central charge for fixed $\gamma/J = 0.7$ and varying values of
      $\varphi$. The effect of increasing $\varphi$ in the range from 0 to
      $\pi/2$ is qualitatively similar to that of decreasing $\gamma/J$: In the
      behavior of the entanglement entropy on large scales, there is an apparent
      transition from area-law scaling to logarithmic growth. Concomitantly, the
      position of the maximum of the effective central charge shifts to the
      right.}
    \label{fig:entropies_fixedgamma}
\end{figure}

Each of the preceding figures has shown data for a fixed value of $\varphi$ but
varying values of the measurement rate $\gamma/J$. This is sufficient to
demonstrate that changing either $\gamma/J$ or $\varphi$ does not induce an
entanglement transition; instead, changing these parameters only modifies the
length scales delimiting the critical and area-law regimes. Nevertheless, it is
interesting to ask how the effect of changing $\gamma/J$ compares to that of
changing $\varphi$. As we have noted already in Sec.~\ref{sec:NLSM-action-RG},
the scale $l_{\varphi, *}$, beyond which area-law entanglement is expected,
depends exponentially on $J/[\gamma \cos(\varphi)]$, meaning that increasing
$\varphi$ toward $\pi/2$ amounts to decreasing the value of
$\gamma/J$. Qualitatively, this relation between $\varphi$ and $\gamma/J$ is
compatible with the observation of an unraveling-induced entanglement transition
in a similar model in Ref.~\cite{Eissler2025}. Furthermore, it is supported by
the data shown in Fig.~\ref{fig:entropies_fixedgamma}. In particular,
Fig.~\ref{fig:entropies_fixedgamma}(a) shows the entanglement entropy at a fixed
value of $\gamma/J = 0.7$ and different values of $\varphi$. Increasing $\varphi$
appears to induce a transition from area-law scaling to logarithmic growth of
the entanglement entropy, similarly to the change induced by increasing
$\gamma/J$ illustrated in
Fig.~\ref{fig:entanglement-entropy-central-charge}(a--c). Nonetheless, as
Fig.~\ref{fig:entropies_fixedgamma}(b) shows, the scale-dependent effective
central charge falls off on large scales, indicating that logarithmic growth of
the entanglement entropy is only transient. We note, however, that including the
unraveling phase in Eq.~\eqref{eq:sse} results not only in a multiplicative
renormalization of the measurement rate, replacing $\gamma$ by
$\gamma \cos(\varphi)$. This becomes clear, for example, from the modified
scaling of $l_c$ and $l_m$ for $\varphi = 5 \pi/12$, shown in the inset in
Fig.~\ref{fig:Clg0_vs_l_plot}(c) and
Fig.~\ref{fig:entanglement-entropy-central-charge}(f), respectively.

\subsection{Unitary unraveling}
\label{sec:unitary-unraveling}

\begin{figure}
  \centering
  \includegraphics{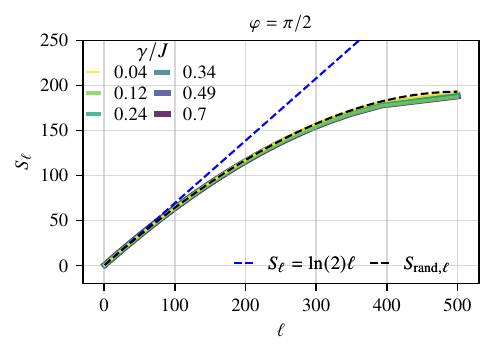}
  \caption{Entanglement entropy density for $\varphi = \pi/2$. For any value of
    $\gamma/J$, the data are consistent with Eq.~\eqref{eq:S-rand} for random
    Gaussian states (black dashed line). We observe deviations for large
    subsystem sizes $\ell$, which are caused by our simulations not having fully
    reached the stationary regime. For small $\ell$, the data exhibit volume-law
    scaling (blue dashed line), as also found for $\varphi \neq \pi/2$ in
    Fig.~\ref{fig:entanglement-entropy-central-charge}(a--c).}
  \label{fig:random-states}
\end{figure}

Finally, we briefly discuss the unitary unraveling at $\varphi = \pi/2$.
Numerical results for this case were reported in Refs.~\cite{Cao2019,
  Eissler2025}, showing that the entanglement entropy grows as
$S_{\ell} \sim \sqrt{t/\gamma}$ and saturates at a value that exhibits
volume-law scaling. Interestingly, the saturation value agrees quantitatively
with the time average of the entanglement entropy in the deterministic dynamics
at $\gamma = 0$. Our numerical simulations for $\varphi = \pi/2$ further suggest
that trajectories at late times approach essentially random Gaussian states. We
support this claim with two observations: First, we found that the steady-state
density correlation function is consistent with that of a random Gaussian state,
which for $N, L \to \infty$ with $N/L = 1/2$ is
\begin{equation}
  C_{l - l'} \sim \frac{1}{4} \left( \delta_{l, l'} - \frac{1}{L} \right).
\end{equation}
This follows by inserting Eq.~\eqref{eq:single-particle-density-matrix} into
Eq.~\eqref{eq:connected-density-correlation-function} and averaging over random
isometries $U(t)$~\footnote{To compute the average, we extend $U(t)$ to an
  $L \times L$ unitary matrix by adding $L - N$ columns; this does not affect
  the single-particle density matrix in
  Eq.~\eqref{eq:single-particle-density-matrix}. The subsequent unitary average
  can be carried out using Weingarten calculus~\cite{Collins2006}.}.  Second, as
shown in Fig.~\ref{fig:random-states}, for any value of $\gamma/J$, the
entanglement entropy follows the behavior of random Gaussian states given
by~\cite{Lydzba2020, Lydzba2021}
\begin{equation}
  \label{eq:S-rand}
  \frac{S_{\mathrm{rand}, \ell}}{L} = \left[ \ln(2) - 1 \right] \frac{\ell}{L} + \left( 1 -
    \frac{\ell}{L} \right) \ln \! \left( 1 - \frac{\ell}{L} \right).
\end{equation}
This form is obtained by averaging over Gaussian states with fixed particle
numbers $N \in \{0, \dotsc, L\}$, allowing $N$ to vary across states. By
contrast, in our setting the particle number is fixed at $N = L/2$, yet we still
observe good agreement. We attribute the deviations at $\ell$ close to $L/2$ to
our simulations not having fully reached the stationary regime. Because of the
slow growth $S_{\ell} \sim \sqrt{t/\gamma}$ toward volume-law behavior, reaching
the steady state requires substantially more time for $\varphi = \pi/ 2$ than
for $0 \leq \varphi < \pi/2$. Note that Eq.~\eqref{eq:S-rand} implies volume-law
behavior, $S_{\mathrm{rand}, \ell} \sim \ell$, for any fixed value of the ratio
$\ell/L$~\cite{Cao2019, Eissler2025}. However, in Fig.~\ref{fig:random-states}
we keep the system size $L$ fixed while varying the subsystem size $\ell$. Then,
$S_{\mathrm{rand}, \ell} \sim \ell$ only for $\ell \ll L$.

It is an interesting question how such an ensemble of random Gaussian states is
described analytically by an NLSM. We do not provide a full answer to that
question here, but make the following observations: For $\varphi = \pi/2$, the
matrix $X_{\varphi}$ defined in Eq.~\eqref{eq:X-phi} reduces to the imaginary
unit, such that Eq.~\eqref{eq:L-M-G} simplifies to
\begin{equation}
  \label{eq:L-M-G-phi=pi/2}
  \imag \mathcal{L}_M[\mathcal{G}] = - \frac{R}{4} - \frac{1}{2} \left[ \tr \!
    \left( \mathcal{G}^2 \right) - \tr \! \left( \mathcal{G} \right)^2 \right].
\end{equation}
Thus, unlike the case $\varphi < \pi/2$, there is no distinction in symmetry
between $R=1$ and $R>1$ for $\varphi = \pi/2$: Both Eqs.~\eqref{eq:L-M-G-R=1}
and~\eqref{eq:L-M-G-phi=pi/2} are invariant under arbitrary transformations
$\mathcal{G} \mapsto \mathcal{R} \mathcal{G} \mathcal{R}^{-1}$. Consequently, in
the construction of the NLSM manifold, the restriction in
Eq.~\eqref{eq:R-prime-sigma-z} does not apply, and the manifold for
$\varphi = \pi/2$ is the complex Grassmannian
$\mathrm{U}(2R)/\mathrm{U}(R)\times\mathrm{U}(R)$. Another important difference
from the case $\varphi < \pi/2$, with NLSM Lagrangian given in
Eq.~\eqref{eq:L-replicon}, is that the temporal derivative for $\varphi = \pi/2$
is presumably first order. This contrasts with the NLSMs appearing in the theory
of 2D disordered systems, where time plays the role of a second spatial
dimension and both directions enter on equal footing.

\section{Conclusions}
\label{sec:conclusions}

In this work, we have analyzed continuously monitored one-dimensional free
fermions in the presence of an unraveling phase $\varphi$, which interpolates
between conventional quantum state diffusion and a unitary unraveling
corresponding to classical noise. Our main finding is that neither varying the
measurement rate $\gamma$ nor tuning the unraveling phase $\varphi$ can drive a
genuine entanglement transition. Instead, the system ultimately exhibits
area-law entanglement beyond the exponentially large scale
$l_{\varphi, *}$~\eqref{eq:l-phi-star}. Apparent signatures of a critical phase,
including logarithmic growth of entanglement and algebraic decay of
correlations, persist only below a crossover scale that grows algebraically with
$J/\gamma$. Our results thus indicate that the apparent KT transition in related
models is a finite-size crossover phenomenon~\cite{Alberton2021, Eissler2025}.

Nevertheless, varying the unraveling phase $\varphi$ affects the dynamics in
nontrivial ways: First, for $\varphi = 0$, our model belongs to the chiral
orthogonal symmetry class BDI; instead, for $\varphi \neq 0$, it belongs to the
chiral unitary class AIII. This distinction is reflected in the universal slope
of the weak-localization correction. Second, increasing $\varphi$ in the range
$0 \leq \varphi < \pi/2$ has a qualitatively similar effect to decreasing
$\gamma$ at fixed $\varphi$, in line with the observation of an
unraveling-induced transitions reported previously~\cite{Eissler2025}. Third,
when $\varphi$ is increased beyond a certain value between $\pi/4$ and $\pi/3$,
novel nonuniversal features appear at short scales, including a maximum of the
scale-dependent effective central charge at $l_m \sim \gamma^{-1}$, in contrast
to the scaling $l_m \sim \gamma^{-3/2}$ that we observe for smaller values of
$\varphi$. These features affect also the scaling of the upper boundary of the
critical range, which is modified from $l_c \sim \gamma^{-2}$ to
$l_c \sim \gamma^{-2/3}$. However, these modifications due to nonuniversal
features are a consequence of our chosen definition of $l_c$; we find the same
\emph{universal} behavior, as captured by the weak-localization correction, for
all values of the unraveling phase in the range $0 \leq \varphi < \pi/2$, and
the weak-localization phenomenon implies the absence of an entanglement
transition. Only the unitary unravelling at $\varphi = \pi/2$ leads to
qualitatively distinct universal behavior. Its description in terms of an NLSM
is an interesting subject for future studies.

Our findings indicate that, for free fermionic systems, the choice of
unraveling---whether formulated in terms of a quantum jump process or quantum
state diffusion, and including further modifications such as the unraveling
phase considered here---does not determine the presence or absence of a
measurement-induced phase transition. Rather, the absence of such a transition
in various models of 1D free fermionic systems with conserved particle number
can be traced to a common $\mathrm{SU}(R)$ symmetry. As noted in
Sec.~\ref{sec:target-manifold}, conservation of particle number within the 1D
fermionic chain is, in fact, not required for this symmetry to arise. It also
persists when the chain is coupled to reservoirs with which it can exchange
particles, provided that the total particle number of the combined system,
consisting of the chain and reservoirs, remains conserved~\cite{Starchl2025}. An
important question for future research is whether this insensitivity to the
choice of unraveling can be established in full generality. It would also be
worthwhile to extend the analysis of different unravelings to observables beyond
the local occupation numbers considered here. Alternative observables may give
rise to qualitatively distinct dynamical behavior. For example, bilocal
observables have been shown to induce superdiffusive scaling of the entanglement
entropy~\cite{Poboiko2025a}. How such phenomena depend on the choice of
unraveling remains an open question.

Looking further ahead, it will be interesting to explore whether the critical
phase can be stabilized by relaxing the constraints of the present model, for
instance, by introducing additional ingredients such as
disorder~\cite{Szyniszewski2023, Szyniszewski24, Matsubara2025, Qiu2025,
  Zhao2025}, Floquet driving~\cite{Chatterjee2025}, or nonlocal measurements. An
intriguing question is whether the singular point $\varphi = \pi/2$ in our model
could, through some mechanism, be extended into a stable phase of nonequilibrium
quantum matter that occupies a finite region of parameter space. Finally,
measurement-induced transitions are known to occur in fermionic systems in
higher dimensions~\cite{Poboiko2024, Chahine2024}, in the presence of
interactions~\cite{Poboiko2025, Guo2025}, or in Majorana models with broken
particle-number conservation~\cite{Fava2023}. Elucidating the role of different
unravelings in these settings constitutes an important direction for future
work.

\begin{acknowledgments}
  We thank Hannes Pichler for interesting discussions. T.V., E.S., and L.S.\
  acknowledge support from the Austrian Science Fund (FWF) through the project
  10.55776/COE1, and from the European Union - NextGenerationEU.
\end{acknowledgments}

\subsection*{Data availability}

The data that support the findings of this article are openly
available~\cite{Niederegger2025}.

\appendix

\section{Derivation of the \sse{} for general quadrature measurements}
\label{sec:sse-derivation}

In this appendix, we provide a formal derivation of the \sse~\eqref{eq:sse} and
the corresponding linear \sse, rewritten in Eq.~\eqref{eq:linear-sme} as a
discrete-time linear stochastic master equation. Our approach is not tied to a
microscopic model of a specific physical realization; instead, it is intended to
clarify the conceptual role of the unraveling phase $\varphi$. Along the way, we
also highlight connections to alternative formulations of continuous
measurements used in Refs.~\cite{Chahine2024, Fava2023, Fava2024}.

\subsection{Setup}

We consider a quantum system coupled to a bosonic ancilla on which measurements
are performed~\cite{Gardiner2015}. Our aim is to demonstrate that the phase
factor $\e^{\imag \varphi}$ in Eq.~\eqref{eq:sse} arises naturally from measuring
a general quadrature of the ancilla. The derivation of the stochastic
Schr\"odinger equation~\eqref{eq:sse}, however, is fully general, and we
therefore leave the nature of the system unspecified at this stage.

We derive the evolution equation~\eqref{eq:sse} by considering the dynamics over
a finite time step $\Delta t$ and subsequently taking the limit
$\Delta t \to \diff t$. The ancilla, which is described by bosonic annihilation
and creation operators $\hat{b}$ and $\hat{b}^{\dagger}$, respectively, is
initialized in the vacuum state, $\hat{b} \ket{0} = 0$. In the usual quantum
optics treatment, $\hat{b}$ and $\hat{b}^{\dagger}$ are quantum It\^o
increments~\cite{Gardiner2015, Wiseman2010}. Accordingly, their commutator is
determined by the step size,
\begin{equation}
  \label{eq:b-commutator-Delta-t}
  \left[ \hat{b}, \hat{b}^{\dagger} \right] = \Delta t.
\end{equation}
As stated above, we leave the nature of the system unspecified. We denote the
initial state of the system by $\ket{\psi_0}$. The initial state of system and
ancilla is thus
\begin{equation}
  \label{eq:system-ancilla-initial-state}
  \ket{\Psi_0} = \ket{\psi_0} \otimes \ket{0}.
\end{equation}
During the time step, the system and ancilla are entangled by applying the
unitary
\begin{equation}
  \label{eq:U-c-Hermitian-b}
  \hat{U} = \e^{\sqrt{\gamma} \hat{c} \left( \hat{b}^{\dagger} - \hat{b}\right)},
\end{equation}
where $\hat{c} = \hat{c}^{\dagger}$ is a system observable. To illustrate the
effect of the entangling operation~\eqref{eq:U-c-Hermitian-b}, consider the simple case
that the system is prepared in an eigenstate of $\hat{c}$. Then, applying
Eq.~\eqref{eq:U-c-Hermitian-b} to the initial
state~\eqref{eq:system-ancilla-initial-state} displaces the bosonic vacuum. The
displacement has magnitude
$\langle \hat{c}_0 \rangle = \langle \psi_0 | \hat{c} | \psi_0 \rangle$ and is
directed along the quadrature
$\hat{x} = (\hat{b} + \hat{b}^{\dagger}) / \sqrt{2}$. Measuring this quadrature
thus provides direct information about $\langle \hat{c}_0 \rangle$.

\subsection{Equivalence between general quadrature measurements and coupling to
  non-Hermitian system operators}

We now generalize the scenario above by either measuring a more general
quadrature or, equivalently, by using a non-Hermitian operator $\hat{c}$,
obtained by multiplying a Hermitian operator by a complex phase factor. Let us
begin by showing that these two approaches are indeed equivalent. To this end,
we define the rotated field quadratures
\begin{equation}
  \hat{x}_{\varphi} = \frac{1}{\sqrt{2}} \left( \hat{b} \e^{\imag \varphi} +
    \hat{b}^{\dagger} \e^{- \imag \varphi} \right), \quad \hat{p}_{\varphi} =
  \frac{1}{\sqrt{2} \imag} \left( \hat{b} \e^{\imag \varphi} -
    \hat{b}^{\dagger} \e^{- \imag \varphi} \right),
\end{equation}
which obey the commutation relation
\begin{equation}
  \label{eq:x-phi-p-phi-commutator}
  \left[ \hat{x}_{\varphi}, \hat{p}_{\varphi} \right] = \imag \Delta t.
\end{equation}
Quadratures with different values of $\varphi$ are unitarily related:
\begin{equation}
  \label{eq:x-phi-x-0}
  \hat{x}_{\varphi} = \hat{R}_{\varphi}^{\dagger} \hat{x}
  \hat{R}_{\varphi}, \qquad \hat{R}_{\varphi} = \e^{\imag \varphi
    \hat{b}^{\dagger} \hat{b}}.
\end{equation}

After the entangling operation Eq.~\eqref{eq:U-c-Hermitian-b}, we perform a
measurement of $\hat{x}_{\varphi}$ on the ancilla. We denote the outcome of the
measurement by $x_m$. The measurement projects the ancilla onto the
corresponding eigenstate $\ket{x_{\varphi} = x_m}$ of $\hat{x}_{\varphi}$; the
effect of the measurement on the system can be expressed through the Kraus
operator $\hat{K}_{x_m}$ that is defined by
\begin{equation}
  \label{eq:K-x-0}
  \hat{K}_{x_m} \ket{\psi_0} = \langle x_{\varphi} = x_m | \hat{U} | \Psi_0 \rangle \,\, \Leftrightarrow
  \,\, \hat{K}_{x_m} = \langle x_{\varphi} = x_m | \hat{U} | 0 \rangle.
\end{equation}
From Eq.~\eqref{eq:x-phi-x-0} it follows that
$\ket{x_{\varphi} = x_m} = \hat{R}_{\varphi}^{\dagger} \ket{x = x_m}$; that is,
eigenvectors of $\hat{x}_{\varphi}$ and $\hat{x}$ are related by
$\hat{R}_{\varphi}^{\dagger}$. Using this and the fact that the transformation
$\hat{R}_{\varphi}$ leaves the bosonic vacuum invariant,
$\hat{R}_{\varphi} \ket{0} = \ket{0}$, we obtain
\begin{equation}
  \label{eq:K-x-phi}
  \hat{K}_{x_m} = \langle x = x_m | \hat{U}_{\varphi} | 0 \rangle,
\end{equation}
where
\begin{equation}
  \label{eq:U-c-phi-b}
  \hat{U}_{\varphi} = \hat{R}_{\varphi} \hat{U} \hat{R}_{\varphi}^{\dagger} =
  \e^{\sqrt{\gamma} \left( \hat{c}_{\varphi} \hat{b}^{\dagger} -
      \hat{c}_{\varphi}^{\dagger} \hat{b} \right)}, \qquad \hat{c}_{\varphi} =
  \hat{c} \e^{\imag \varphi}.
\end{equation}
Compare Eqs.~\eqref{eq:K-x-0} and \eqref{eq:K-x-phi}: The former describes the
measurement of $\hat{x}_{\varphi}$ after coupling the system to the bath via
Eq.~\eqref{eq:U-c-Hermitian-b}, with a Hermitian system operator $\hat{c}$; the
latter describes the measurement of $\hat{x}$ after coupling the system to the
bath via Eq.~\eqref{eq:U-c-phi-b}, where the phase factor $\e^{\imag \varphi}$
has been absorbed in the definition of the non-Hermitian system operator
$\hat{c}_{\varphi}$. Both equations provide equivalent definitions of the Kraus
operator. Therefore, without loss of generality, we can restrict ourselves to
measurements of $\hat{x}$ while allowing for an entangling operation with a
non-Hermitian system operator $\hat{c} \neq \hat{c}^{\dagger}$. We thus consider
\begin{equation}
  \hat{K}_x = \braket{x | \hat{U} | 0},
\end{equation}
with
\begin{equation}
  \label{eq:U-c-b}
  \hat{U} = \e^{\sqrt{\gamma} \left( \hat{c} \hat{b}^{\dagger} -
      \hat{c}^{\dagger} \hat{b} \right)}.
\end{equation}
Here and in the following, we omit the subscript $m$ of the measurement
outcome.

\subsection{Kraus operators}

We now aim to determine the explicit form of the Kraus operators $\hat{K}_x$.
As discussed above, from the ancilla's perspective, $\hat{U}$ in
Eq.~\eqref{eq:U-c-b} essentially acts as a displacement operator, with the
complication that $\hat{c}$ and $\hat{c}^\dagger$ may not commute. This
noncommutativity, however, affects only higher-order terms in $\Delta t$. To
demonstrate this, one can, for instance, expand $\hat{U}$ in $\Delta t$, as done
in the derivation of the stochastic Schrödinger equation in
Ref.~\cite{Gardiner2015}. Here, we instead follow the alternative approach of
factorizing $\hat{U}$ using the Zassenhaus
formula~\cite{Wilcox1967}. Equation~\eqref{eq:b-commutator-Delta-t} implies
$\hat{b} \sim \sqrt{\Delta t}$. Therefore, to leading order in $\Delta t$, we
obtain
\begin{equation}
  \hat{U} = \e^{\sqrt{\gamma} \hat{c} \hat{b}^{\dagger}} \e^{- \sqrt{\gamma}
    \hat{c}^{\dagger} \hat{b}} \e^{\frac{\gamma}{2} \hat{c} \hat{c}^{\dagger}
    \hat{b}^{\dagger} \hat{b}} \e^{- \frac{\gamma}{2} \hat{c}^{\dagger} \hat{c}
    \hat{b} \hat{b}^{\dagger}}. 
\end{equation}
Then, using $\hat{b} \hat{b}^{\dagger} \ket{0} = \left[ \hat{b},
  \hat{b}^{\dagger} \right] \ket{0} = \Delta t \ket{0}$, we find
\begin{equation}
  \hat{K}_x = M_x(\hat{c}) \e^{- \frac{\gamma \Delta t}{2}
    \hat{c}^{\dagger} \hat{c}}, \quad M_x(\hat{c}) = \braket{x |
    \e^{\sqrt{\gamma} \hat{c} \hat{b}^{\dagger}} | 0}.
\end{equation}
To evaluate $M_x(\hat{c})$, we can treat $\hat{c}$, which is the only system
operator appearing in this expression, as a number $c \in \C$. We introduce a
displacement operator $\hat{V}$,
\begin{equation}
  \hat{V} = \e^{\sqrt{\gamma} \left( c \hat{b}^{\dagger} - c^{*} \hat{b}
    \right)} = \e^{\sqrt{\gamma} c \hat{b}^{\dagger}} \e^{- \sqrt{\gamma} c^{*}
    \hat{b}} \e^{- \frac{\gamma \Delta t}{2} \abs{c}^2},
\end{equation}
in terms of which we can write
\begin{equation}
  M_x(c) = \e^{\frac{\gamma \Delta t}{2} \abs{c}^2} \braket{x | \hat{V}
    | 0} = \frac{1}{\left( \pi \Delta t \right)^{1/4}} \e^{- \frac{1}{2
      \Delta t} \left( x - \sqrt{2 \gamma} \Delta t c \right)^2 + \frac{\gamma
      \Delta t}{2} c^2}.
\end{equation}
Note that the representation of the coherent state $\hat{V} \ket{0}$ in the
basis $\ket{x}$ is modified as compared to the usual one due to the factor
$\Delta t$ in the commutation relations in Eqs.~\eqref{eq:b-commutator-Delta-t}
and~\eqref{eq:x-phi-p-phi-commutator}. Replacing $c$ by the operator $\hat{c}$
yields the desired explicit form of the Kraus operators:
\begin{equation}
  \label{eq:K-x}
  \hat{K}_x = \frac{1}{\left( \pi \Delta t \right)^{1/4}} \e^{- \frac{x^2}{2
      \Delta t} + \sqrt{2 \gamma} x \hat{c} - \frac{\gamma \Delta t}{2} \left(
      \hat{c} + \hat{c}^{\dagger} \right) \hat{c}},
\end{equation}
where we again keep only the leading order in $\Delta t$, taking into account
that $x \sim \sqrt{\Delta t}$ due to the first term in the exponent. The Kraus
operators obey the completeness relation
\begin{equation}
  \int_{-\infty}^{\infty} \diff x \, \hat{K}_x^{\dagger} \hat{K}_x = 1.
\end{equation}

\subsection{Quantum trajectories}

The formulation of the measurement process in terms of the Kraus
operators~\eqref{eq:K-x} provides the basis for defining quantum
trajectories. We iterate the measurement process described above $n$ times,
corresponding to a time evolution of total duration $T = n \Delta t$. After each
measurement, the ancilla is reinitialized in the vacuum state. We denote the
resulting sequence of measurement outcomes by
$\mathbf{x} = \left( x_0, \dotsc, x_{n - 1} \right)$. According to Born's rule,
the probability $P_{\mathbf{x}}$ of observing this sequence of outcomes is given
by the squared norm of the resulting state,
\begin{equation}
  \label{eq:u-n}
  \ket{u_n} = \hat{K}_{x_{n - 1}} \dotsb \hat{K}_{x_0} \ket{\psi_0},
\end{equation}
where we leave the dependence of the state on the measurement outcomes implicit;
that is,
\begin{equation}
  \label{eq:P-x-u-n}
  P_{\mathbf{x}} = \norm{\ket{u_n}}^2.
\end{equation}
The average of a function of the normalized state,
$\ket{\psi_n} = \ket{u_n}/\norm{\ket{u_n}}$, can be calculated as
\begin{equation}
  \label{eq:average-f-psi}
  \overline{f(\ket{\psi_n})} = \int \diff \mathbf{x} \,
  P_{\mathbf{x}} f(\ket{\psi_n}).
\end{equation}
Important examples include the connected density correlation
function~\eqref{eq:C-l-l-prime-t} and the entanglement
entropy~\eqref{eq:entanglement-entropy}.

Equations~\eqref{eq:u-n} and~\eqref{eq:P-x-u-n} provide a complete description
of the system dynamics under repeated measurements. We proceed to rephrase this
description, first as a linear evolution equation for an unnormalized state, and
further below as a nonlinear evolution equation for a normalized state.

\subsection{Linear \sse}

Instead of labeling measurement outcomes by the quadrature $x$, we use the
variable $\Delta W = \sqrt{2} x$ in the following. We define the ostensible
distribution of $\Delta W$ as~\cite{Wiseman2010}
\begin{equation}
  \label{eq:Q-Delta-W-l-n}
  Q_{\Delta W} = \frac{1}{\sqrt{2 \pi \Delta t}} \exp \! \left( - \frac{\Delta
      W^2}{2 \Delta t} \right),
\end{equation}
and we factor out the ostensible distribution from Eq.~\eqref{eq:K-x} to define
a new set of Kraus operators:
\begin{equation}
  \label{eq:J-Kraus-operators}
  \hat{J}_{\Delta W} = \left. \frac{1}{\sqrt{Q_{\Delta W}}} \frac{\hat{K}_x}{2^{1/4}} \right\rvert_{x =
    \Delta W/\sqrt{2}} = \e^{\sqrt{\gamma} \Delta W \hat{c} - \frac{\gamma \Delta t}{2} \left( \hat{c}
      + \hat{c}^{\dagger} \right) \hat{c}}.
\end{equation}
These operators obey the completeness relation
\begin{equation}
  \int_{-\infty}^{\infty} \diff \Delta W \, Q_{\Delta W} \hat{J}_{\Delta
    W}^{\dagger} \hat{J}_{\Delta W} = 1. 
\end{equation}
Consider now a trajectory, defined as in Eq.~\eqref{eq:u-n} but with the Kraus
operators $\hat{K}_x$ replaced by $\hat{J}_{\Delta W}$:
\begin{equation}
  \label{eq:v-n}
  \ket{v_n} = \hat{J}_{\Delta W_{n - 1}} \dotsb \hat{J}_{\Delta
    W_0} \ket{\psi_0}.
\end{equation}
As above, we leave the dependence of the state on the sequence of measurement
outcomes
$\Delta \mathbf{W} = \left( \Delta W_0, \dotsc, \Delta W_{n - 1} \right)$
implicit. We denote the norm of the state by
\begin{equation}
  R_{\Delta \mathbf{W}} = \norm{\ket{v_n}}^2 = \left. P_{\mathbf{x}} \middle/
    \left( 2^{n/2} Q_{\Delta \mathbf{W}} \right) \right\rvert_{\mathbf{x} =
    \Delta \mathbf{W}/\sqrt{2}},
\end{equation}
where the ostensible probability of a the full sequence of measurement outcomes
is the product of probabilities of the outcomes of individual measurements:
\begin{equation}
  Q_{\Delta \mathbf{W}} = \prod_{n' = 0}^{n - 1} Q_{\Delta W_{n'}}.
\end{equation}
However, the \emph{true} probability of a trajectory is the product of the
ostensible probability $Q_{\Delta \mathbf{W}}$ and the norm of the state
$R_{\Delta \mathbf{W}}$. Therefore, the average of a function of the normalized
state $\ket{\psi_n} = \ket{v_n}/\norm{\ket{v_n}}$ should be computed as
\begin{equation}
  \label{eq:f-average-Q-P-tilde}  
  \overline{f(\ket{\psi_n})} = \int \diff \Delta \mathbf{W}
  \, Q_{\Delta \mathbf{W}} R_{\Delta \mathbf{W}} f(\ket{\psi_n}) = \int \diff \mathbf{x} \, P_{\mathbf{x}}
  f(\ket{\psi_n}),  
\end{equation}
which correctly reproduces Eq.~\eqref{eq:average-f-psi}.

The formulation of the stochastic dynamics in terms of the Kraus operators
$\hat{J}_{\Delta W}$ leads to the usual form of the linear
\sse~\cite{Jacobs2006, Jacobs2014}. To see this, we expand
Eq.~\eqref{eq:J-Kraus-operators} up to first order in $\Delta t$, using
$\Delta W^2 = \Delta t$~\cite{Jacobs2006},
\begin{equation}
  \label{eq:J-Delta-W-expansion}
  \hat{J}_{\Delta W} \sim 1 - \frac{\gamma}{2} \hat{c}^{\dagger}
  \hat{c} \Delta t + \sqrt{\gamma} \hat{c} \Delta W.
\end{equation}
We can generalize the Kraus operators to include a Hamiltonian contribution that
acts on the system alone: This does not affect the above derivation. The
resulting linear evolution equation for the unnormalized state~\eqref{eq:v-n}
reads
\begin{equation}
  \label{eq:linear-sse-c}
  \ket{v_{n + 1}} = \left( 1 - \imag \hat{H} \Delta t - \frac{\gamma}{2}
    \hat{c}^{\dagger} \hat{c} \Delta t + \sqrt{\gamma} \hat{c} \Delta W_n
  \right) \ket{v_n},
\end{equation}
and the corresponding linear master equation for the unnormalized density matrix
$\hat{D}_n = \ket{v_n} \bra{v_n}$ is given by
\begin{multline}
  \label{eq:linear-MEQ-c}
  \hat{D}_{n + 1} = \hat{D}_n - \imag \left[ \hat{H}, \hat{D}_n \right] + \gamma
  \left( \hat{c} \hat{D}_n \hat{c}^{\dagger} - \frac{1}{2} \left\{
      \hat{c}^{\dagger} \hat{c}, \hat{D}_n \right\} \right) \Delta t \\ +
  \sqrt{\gamma} \left( \hat{c} \hat{D}_n + \hat{D}_n \hat{c}^{\dagger} \right)
  \Delta W_n.
\end{multline}
Finally, we specialize this equation to the fermionic lattice system considered
in the main text. This entails performing, during each time step of duration
$\Delta t$, measurements on all lattice sites with
$\hat{c} = \hat{n}_l \e^{\imag \varphi}$, rather than a single measurement. To
leading order in $\Delta t$, the measurement outcomes satisfy
$\Delta W_{l,n} \Delta W_{l',n} = \delta_{l,l'} \Delta t$, yielding
Eq.~\eqref{eq:linear-sme}. An alternative derivation of the stochastic master
equation, based on general results from stochastic process theory, is given in
Ref.~\cite{Fava2024}.

\subsection{Other formulations of linear stochastic dynamics}
\label{sec:linear-others}

We find it useful to briefly digress and discuss the relation between our
approach and other formulations of linear stochastic dynamics that have been
employed to derive replica field theories for Hermitian observables,
$\hat{c} = \hat{c}^{\dagger}$~\cite{Chahine2024, Fava2023, Fava2024}. To this
end, we first introduce the parameter
\begin{equation}
  \label{eq:alpha-q}
  \alpha = \frac{x}{\sqrt{2 \gamma} \Delta t},
\end{equation}
and define new Kraus operators
\begin{equation}
  \label{eq:K-alpha}
  \hat{K}_{\alpha} = \left. \left( 2 \gamma \Delta t^2 \right)^{1/4} \hat{K}_x
  \right\rvert_{x = \sqrt{2 \gamma} \Delta t} = \left( \frac{2 \gamma \Delta t}{\pi} \right)^{1/4} \e^{-
    \gamma \Delta t \left( \alpha - \hat{c} \right)^2}.
\end{equation}
A formulation of continuous measurements using Kraus operators of the form
Eq.~\eqref{eq:K-alpha} was employed in Ref.~\cite{Chahine2024} to derive a
replica master equation, which was then rewritten as a Keldysh field theory. A
slightly different but equivalent formulation is used in Refs.~\cite{Fava2023,
  Fava2024}: As in Eq.~\eqref{eq:J-Kraus-operators}, these works factor out the
ostensible distribution of $\alpha$,
\begin{equation}
  p_{\alpha} = \sqrt{\frac{2 \gamma \Delta t}{\pi}} \e^{- 2 \gamma \Delta t \alpha^2}.
\end{equation}

\subsection{Nonlinear \sse}
\label{sec:SSE}

We now return to the case of non-Hermitian operators,
$\hat{c} \neq \hat{c}^{\dagger}$, and derive the nonlinear stochastic
Schrödinger equation for the normalized state. This follows a standard
procedure~\cite{Gardiner2015}, which we summarize here for
completeness. The probability of obtaining the measurement outcome
$\Delta W_n$ in time step $n$ is
\begin{equation}
  \begin{split}
    P_{\Delta W_n} & = Q_{\Delta W_n} \braket{\psi_n | \hat{J}_{\Delta
        W_n}^{\dagger} \hat{J}_{\Delta W_n} | \psi_n} \\ 
    & =
    \frac{1}{\sqrt{2 \pi \Delta t}} \e^{- \frac{1}{2 \Delta t} \left( \Delta W_n
        - \sqrt{\gamma} \langle \hat{a}_n \rangle \Delta t \right)^2},
  \end{split}
\end{equation}
where $\hat{a} = \hat{c} + \hat{c}^{\dagger}$ and
$\langle \hat{a}_n \rangle = \braket{\psi_n | \hat{a} | \psi_n}$. This suggests
shifting
$\Delta W_n \to \Delta W_n + \sqrt{\gamma} \langle \hat{a}_n \rangle \Delta t$,
where the shifted $\Delta W_n$ is a Wiener increment with statistics given by
Eq.~\eqref{eq:Q-Delta-W-l-n}. With this substitution,
Eq.~\eqref{eq:linear-sse-c} becomes
\begin{equation}
  \ket{v_{n + 1}} = \left( 1 - \imag \hat{H} \Delta t - \frac{\gamma}{2}
    \hat{c}^{\dagger} \hat{c} \Delta t + \gamma \langle \hat{a}_n \rangle
    \hat{c} \Delta t + \sqrt{\gamma} \hat{c} \Delta W_n \right) \ket{v_n}.
\end{equation}
The corresponding evolution equation for the normalized state,
$\ket{\psi_n} = \ket{v_n}/\norm{\ket{v_n}}$, reads
\begin{multline}
  \ket{\psi_{n + 1}} = \left[ 1 - \imag \hat{H} \Delta t - \frac{\gamma}{2}
    \left( \hat{c}^{\dagger} \hat{c} - \langle \hat{a}_n \rangle \hat{c} +
      \frac{1}{4} \langle \hat{a}_n \rangle^2 \right) \Delta t \right. \\
  \left. + \sqrt{\gamma} \left( \hat{c} - \tfrac{1}{2} \langle \hat{a}_n \rangle
    \right) \Delta W_n \right] \ket{\psi_n}.
\end{multline}
A time-dependent gauge transformation,
$\ket{\psi_n} \mapsto \e^{-\imag \phi_n} \ket{\psi_n}$ with
$\phi_n = \sum_{n' = 0}^n \Delta \phi_{n'}$ and
\begin{equation}
  \Delta \phi_n = -
  \frac{\gamma}{4} \langle \hat{a}_n \rangle \langle \hat{b}_n \rangle
  \Delta t - \frac{\sqrt{\gamma}}{2} \langle \hat{b}_n \rangle \Delta W_n,
\end{equation}
where $\hat{b} = - \imag \left( \hat{c} - \hat{c}^{\dagger} \right)$, yields
\begin{multline}
  \ket{\psi_{n + 1}} = \left[ 1 - \imag \hat{H} \Delta t - \frac{\gamma}{2}
    \left( \hat{c}^{\dagger} \hat{c} - 2 \hat{c} \langle \hat{c}_n^{\dagger}
      \rangle + \abs{\langle \hat{c}_n \rangle}^2 \right) \Delta t \right. \\
  \left. \vphantom{\frac{1}{2}} + \sqrt{\gamma} \left( \hat{c} - \langle
      \hat{c}_n \rangle \right) \Delta W_n \right] \ket{\psi_n}.
\end{multline}
In the continuous-time limit, $\Delta t \to \diff t$ and
$\Delta W_n \to \diff W(t)$, this reduces to the standard nonlinear stochastic
Schrödinger equation~\cite{Gardiner2015}. To obtain Eq.~\eqref{eq:sse}, which
describes continuous measurements carried out simultaneously on all lattice
sites, we set $\hat{c} = \hat{n}_l \e^{\imag \varphi}$ and sum over $l$.

\section{Numerical solution for the density correlation function}
\label{sec:wienerhopf}

Here, we detail the numerical procedure we use to calculate the density
correlation function. In Eq.~\eqref{eq:C-l-Gaussian-full-numerical}, the density
correlation function is expressed in terms of $\mathcal{C}$, which in turn is
expressed in Eq.~\eqref{eq:C-pi/4-case} in terms of $L$, the solution of the
Wiener-Hopf integral equation~\eqref{eq:Wiener-Hopf}. To find $L$, we rewrite
Eq.~\eqref{eq:Wiener-Hopf} in terms of the dimensionless momentum
$u = 2l_0\sin(q)$ and time $\tau = \gamma t$ as
\begin{equation}
  \tilde{L}_u(\tau,\tau^\prime) -
  \int_{-\infty}^0\mathrm{d}\tau''\,\mathcal{B}_{+,u}(\tau -
  \tau'')\tilde{L}_u(\tau'', \tau')=\delta(\tau-\tau'),
\end{equation}
where $\tilde{L}_u(\tau, \tau')=L_q(t, t')/\gamma$. A substitution of
integration variables $\tau = 1 - 1/s$ maps the range of integration from
$\tau \in (-\infty, 0]$ to $s \in (0, 1]$, leading to
\begin{equation}
  \label{eq:wiener-hopf-cutless}
  \tilde{L}_u(s,s') - \int_0^1\diff s''\, \frac{\mathcal{B}_{+,u}(s -
    s'')}{(s'')^2} \tilde{L}_u(s'',s') = s^2 \delta(s-s').
\end{equation}
This integral equation can be recast as a matrix equation that can be solved
numerically by approximating the integral through a Riemann sum. To that end, we
use the partition of the interval $(0, 1]$ which is given by the sequence
$s_n = n \Delta s$ with $\Delta s = 1/N$ such that $s_0 = 0$ and $s_N = 1$. The
discretized form of Eq.~\eqref{eq:wiener-hopf-cutless} thus reads
\begin{equation}
  \tilde{L}_u - \Delta s \mathcal{B}_{+, u} D^{-1} \tilde{L}_u = D/\Delta s,
\end{equation}
which is written here in terms of matrices with elements
$\tilde{L}_{u, n, n'} = \tilde{L}_u(s_n, s_{n'})$,
$\mathcal{B}_{+, u, n, n'} = \mathcal{B}_{+, u}(s_n - s_{n'})$, and
$D_{n, n'} = s_n^2 \delta_{n, n'}$. Solving this equation for $\tilde{L}_u$
yields
\begin{equation}
  \label{eq:wiener-hopf-solution}
  \tilde{L}_u = \left[ \Delta s \left( 1 - \Delta s \mathcal{B}_{+, u} D^{-1}
    \right) \right]^{-1} D.
\end{equation}
We obtain the density correlation function by discretizing
equations~\eqref{eq:C-l-Gaussian-full-numerical} and~\eqref{eq:C-pi/4-case} in a
similar fashion and inserting~\eqref{eq:wiener-hopf-solution}.

\bibliography{bibliography}

\end{document}